\documentclass[aps,twocolumn,groupedaddress,amsmath,amssymb,superscriptaddress]{revtex4}

\usepackage{amsmath}
\usepackage{amssymb}
\usepackage{graphicx}
\usepackage{bm}
\usepackage{epic}
\usepackage{eepic}
\usepackage{pifont}
\usepackage[latin1]{inputenc}
\usepackage{rotating}
\usepackage{color}
\usepackage{nicefrac}
\usepackage{ulem}
\usepackage{longtable}
\usepackage[caption=false]{subfig}

\begin{document}
\title{Interplay between microdynamics and macrorheology in vesicle suspensions}
\author{Badr Kaoui}
\email{badr.kaoui@uni-bayreuth.de}
\affiliation{Theoretical Physics I, University of Bayreuth, Universit\"{a}tsstrasse 30,
D-95447 Bayreuth, Germany}
\affiliation{Department of Applied Physics, Eindhoven University of Technology, Den Dolech 2, 5612 AZ Eindhoven, The Netherlands}
\author{Ruben J. W. Jonk}
\affiliation{Department of Applied Physics, Eindhoven University of Technology, Den Dolech 2, 5612 AZ Eindhoven, The Netherlands}
\author{Jens Harting}
\affiliation{Department of Applied Physics, Eindhoven University of Technology, Den Dolech 2, 5612 AZ Eindhoven, The Netherlands}
\affiliation{Faculty of Science and Technology, Mesa+ Institute, University of Twente, 7500 AE Enschede, The Netherlands}

\date{\today}
\begin{abstract}
The microscopic dynamics of objects suspended in a fluid determines the
macroscopic rheology of a suspension. For example, 
as shown by Danker and Misbah [Phys. Rev. Lett. {\bf 98}, 088104 (2007)], 
the viscosity of a dilute suspension of fluid-filled vesicles is a non-monotonic function of
the viscosity contrast (the ratio between the viscosities of the internal
encapsulated and the external suspending fluids) and exhibits a minimum at the
critical point of the tank-treading-to-tumbling transition. By
performing numerical simulations, we recover this effect and demonstrate that it
persists for a wide range of vesicle parameters such as the concentration,
membrane deformability, or swelling degree.
We also explain why other numerical and experimental studies lead to contradicting results.
Furthermore, our simulations show that this effect even persists in non-dilute
and confined suspensions, but that it becomes less pronounced at higher
concentrations and for more swollen vesicles. For dense suspensions and for
spherical (circular in 2D) vesicles, the intrinsic viscosity tends to depend
weakly on the viscosity contrast.
\end{abstract}
\maketitle
\section{Introduction}
Rheological properties of complex fluids (e.g. suspensions or emulsions) are
not yet fully understood. Their macroscopic behavior is tightly coupled
in a non-trivial way to the dynamics of their components at the microscale.
Understanding this interplay is of importance for many fundamental and
practical applications. Many constitutive laws have been proposed since the
pioneering works of Einstein~\cite{Einstein1906} and
Batchelor~\cite{Batchelor1970} for the rheology of suspensions, in particular
for rigid particles. Suspensions of fluid-filled deformable objects are a
sub-class of complex fluids for which rheology depends on the deformability of
the suspended particles and on the nature of the fluid they encapsulate. The
most known and studied case is blood. How blood flows results from the
micro-structuration of its components, mainly red blood cells (RBCs). For
example, the F\aa hr\ae us-Lindqvist~\cite{Fahraeus1931} effect in blood
vessels --- the blood viscosity decreases with the vessel width is reduced 
--- is caused by the lateral migration of RBCs towards the center of the
vessel. Another example is the effect first proposed by Danker and
Misbah~\cite{Danker2007} which is observed for dilute suspensions of vesicles:
following from the dynamical state of each vesicle at the microscale, the
shear viscosity of the suspension varies in a non-monotonic way as a function
of the viscosity contrast $\Lambda$ (the ratio between the viscosities of the
suspending and the encapsulated fluids). 

Vesicles undergo mainly two states of motion under shear flow: either
\textit{tank-treading} (the particle assumes a steady angle with the flow
direction, while its membrane undergoes a tank-treading-like motion) or
\textit{tumbling} (the particle rotates around its center of
mass)~\cite{Beaucourt2004}. At lower viscosity contrasts, a particle
tank-treads (TT) and at higher viscosity contrasts it tumbles (TB). One way to
trigger the transition from TT to TB is by solely increasing $\Lambda$ beyond a
threshold $\Lambda_{C}$. At this critical point, the viscosity of the
suspension changes from a decreasing to an increasing function of $\Lambda$.
Danker and Misbah predicted this effect theoretically~\cite{Danker2007} and it
was later confirmed experimentally~\cite{Vitkova2008} and numerically using the boundary integral method~\cite{Ghigliotti2010,Zhao2011,Zhao2013,Thiebaud2013}. A similar trend was also observed for RBCs~\cite{Vitkova2008} and capsules~\cite{Bagchi2010}. However,
recent numerical simulations by Lamura and Gompper~\cite{Lamura2013} (based on
the multi-particle collision dynamics method) did not capture this effect.
Instead, the viscosity is found to be a monotonically increasing function of
the viscosity contrast in the range $1 \leq \Lambda \leq 10$, somehow similar to the rheological behavior of an emulsion \cite{Ghigliotti2010,Kennedy1994}.
In the same range of $\Lambda$, experiments of Kantsler \textit{et al} \cite{Kantsler2008} revealed also a monotonic behavior of the viscosity, but as a purely decreasing function of $\Lambda$. 
Thus, there is an apparent contradiction between different studies regarding the dependency of the vesicle suspension viscosity on the viscosity contrast.

In the present paper we recheck independently for the existence of
the Danker-Misbah effect using an alternative simulation technique based on
the lattice-Boltzmann and the immersed boundary
methods~\cite{Kaoui2011,Kaoui2012}. As in Ref.~\cite{Lamura2013} we consider a
confined geometry and a non-zero Reynolds number. The main observable
is the \textit{intrinsic viscosity} $\eta$ and the main control parameter
is the \textit{viscosity contrast} $\Lambda$. We study how the dependence of
$\eta$ on $\Lambda$ changes when varying
the concentration $\phi$, the capillary number ${\rm Ca}$ (via the membrane
rigidity) and the swelling degree $\Delta$. We capture the non-monotonic
behavior of the viscosity as proposed by Danker and Misbah and find that it
persists even when varying $\phi$, ${\rm Ca}$ or $\Delta$. It only becomes
less pronounced for denser suspensions or for very swollen vesicles, but it
does not show a monotonically increasing/decreasing behavior with $\Lambda$ as reported in
Ref.~\cite{Lamura2013,Kantsler2008}.
We explain this disagreement and provide insight into the origins that lead to this apparent contradiction.
\section{Simulation method} 
We only give a short overview on the algorithm and refer
to our previous articles for more details~\cite{Kaoui2011,Kaoui2012,Kaoui2013}.

\textit{\textbf{Fluid dynamics}} -- The dynamics of the involved fluids is
simulated using the lattice-Boltzmann method
(LBM)~\cite{BenziPhysRep,Raabe2004}. The LBM is based on a discrete version of
Boltzmann's equation and recovers the solutions of the Navier-Stokes equations
in the limit of small Knudsen and Mach numbers. Our implementation combines the
standard nine velocity model in two dimensions (D2Q9) with a single relaxation
time Bathnagar-Gross-Krook collision scheme. The computational domain is a
channel with length $L_{x}$ and height $L_{y}$. At the inlet and outlet
of the channel, periodic boundary conditions are imposed. At the bottom a
mid-grid bounce back no-slip boundary is set. The top no-slip boundary is
translated from left to right with a steady velocity ${u}_{tw} = L_y\gamma$,
where $\gamma$ is the shear rate.

\textit{\textbf{Vesicles}} -- Vesicles are closed lipid membranes. They
encapsulate an internal fluid and are suspended in an external fluid.
Their membrane experiences resistance towards bending and
compressing/stretching deformation modes. This gives rise to a restorative
force which in 2D is given by
\begin{equation}
\textbf{f} (s) = \Bigg[ \kappa (\frac{\partial^2c}{\partial s^2} + \frac{c^3}{2} ) - c\zeta \Bigg] \textbf{n} + \frac{\partial \zeta}{\partial s} \textbf{t}.
\label{eq:membrane}
\end{equation}
Here, $c$ is the local membrane curvature, $\kappa$ is the bending modulus (the
membrane rigidity), and $s$ is the arclength coordinate along the membrane.
$\textbf{n}$ and $\textbf{t}$ are the normal and tangential unit vectors,
respectively. $\zeta$ is the effective tension field that enforces the local
inextensible character of the membrane, which leads to the conservation of the
vesicle perimeter $P$. A detailed derivation of the membrane force can be found
in \cite{Kaoui2008}. In addition, we consider that the fluids, inside and
outside the vesicles, to be incompressible Newtonian fluids. This latter leads
to the conservation of the vesicle enclosed area $A$.

\textit{\textbf{Viscosity contrast}} --- The viscosity contrast $\Lambda$ is
defined as the ratio of the internal to the external fluid viscosities. Here,
we restrict ourselves to $ 1 \leq \Lambda \leq 20$. In order to achieve this
numerically, the LBM relaxation time, that is related to the
viscosity~\cite{Raabe2004}, is adjusted depending on whether a fluid node is
located inside or outside a vesicle using the \textit{even-odd
rule}~\cite{Kaoui2012}.

\textit{\textbf{Vesicle-fluid coupling}} --- We couple the fluid flow and the
vesicle dynamics using the immersed boundary method (IBM)~\cite{Peskin2002}: 
an Eulerian regular fixed mesh represents the fluid, while the
vesicles are modeled as Lagrangian moving meshes. The first step consists in
computing the fluid flow with the LBM, as if the membrane does not exist.
Then, the velocity of each membrane point is computed by interpolation of
the velocities of its surrounding fluid nodes. The membrane is advected,
deformed and adopts a new out-of-equilibrium shape. Afterwards, the
restoring membrane force (Eq.~\ref{eq:membrane}) is evaluated and exerted on
the surrounding fluid. These two steps provide a fluid-structure two-way
coupling causing the motion of the vesicles and the disturbance of the externally
applied flow.

\textit{\textbf{Rheology}} --- The effective viscosity $\eta^*$ of a suspension
--- consisting of the suspending fluid and its suspended vesicles --- under
shear flow is calculated using 
\begin{equation}
\eta ^*(t) = \frac{\langle \sigma_{xy}(t)\rangle}{\gamma},
\label{eq:viscosity}
\end{equation}
where $\langle \sigma_{xy}(t) \rangle$ are the
hydrodynamic stresses averaged on the bounding walls~\cite{Kaoui2011},
\begin{equation}
\langle \sigma_{xy}(t) \rangle = \frac{1}{2L_x} \int_{0}^{L_x} \sigma_{xy}(x,t)dx.
\end{equation}
We introduce the dimensionless quantity, 
\begin{equation}
\eta (t) = \frac{\eta ^*(t)-\eta _0}{\eta _0 \phi},
\label{eq:etat}
\end{equation}
where $\eta _0$ is the viscosity of the external suspending fluid and $\phi$
the concentration of the vesicles. $\eta(t)$ measures the
deviation of $\eta^*$ (the viscosity of the fluid in the presence of the
vesicles) from $\eta _0$ (the viscosity of the fluid in the absence of
vesicles) normalized by the quantity $\eta _0 \phi$. Following Refs.~\cite{Lamura2013,Vitkova2008}, we call this quantity the \textit{intrinsic viscosity},
even though we use it not only in the very dilute limit. Other authors
rather call it the \textit{normalized effective viscosity}~\cite{Thiebaud2013}
or the \textit{normalized suspension viscosity}~\cite{Kantsler2008}. We further
use the average of $\eta(t)$ over time
\begin{equation}
\eta = \langle \eta (t) \rangle = \frac{1}{t_f-t_s} \int_{t_s}^{t_f} \eta (t) dt.
\label{eq:etaI}
\end{equation}
The first $t_s= 3 \times 10^5$ timesteps ($\gamma t_s = 62.49$) are
ignored in order to assure that the system has reached a quasi-steady regime.
Data is then taken until the final timestep $t_f=10^6$ ($\gamma t_f = 208.3$). 
In this time interval $[t_s,t_f]$ the system is
in the quasi-steady regime in all simulations -- independent of the chosen values of the
input parameters. This quasi-steady regime is characterized by fluctuations of
the instantaneous intrinsic viscosity around an average value.

\begin{figure*}
\centering
\subfloat[The dynamical behavior at the microscale]{\label{fig:Figure01a}\includegraphics[angle=0,height=0.31\textwidth]{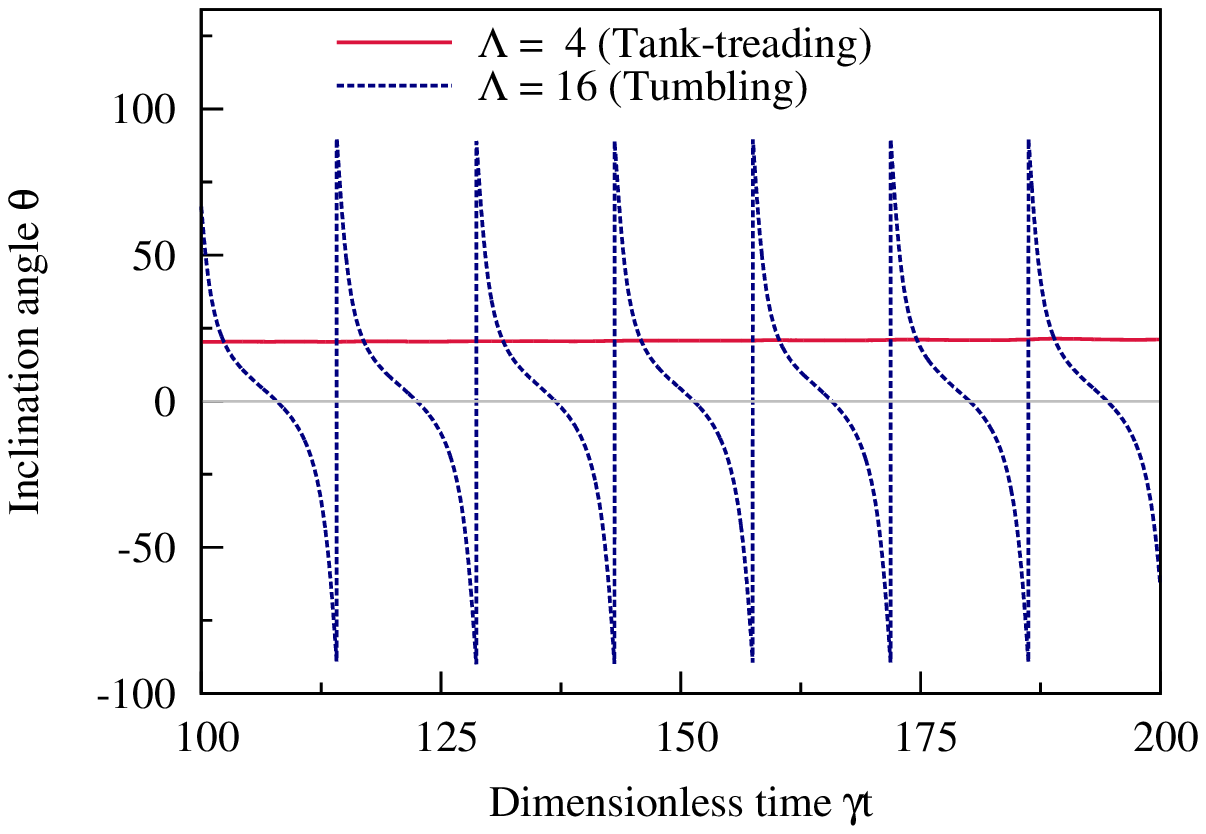}}
\quad\quad\quad 
\subfloat[The rheological behavior at the macroscale]{\label{fig:Figure01b}\includegraphics[angle=0,height=0.31\textwidth]{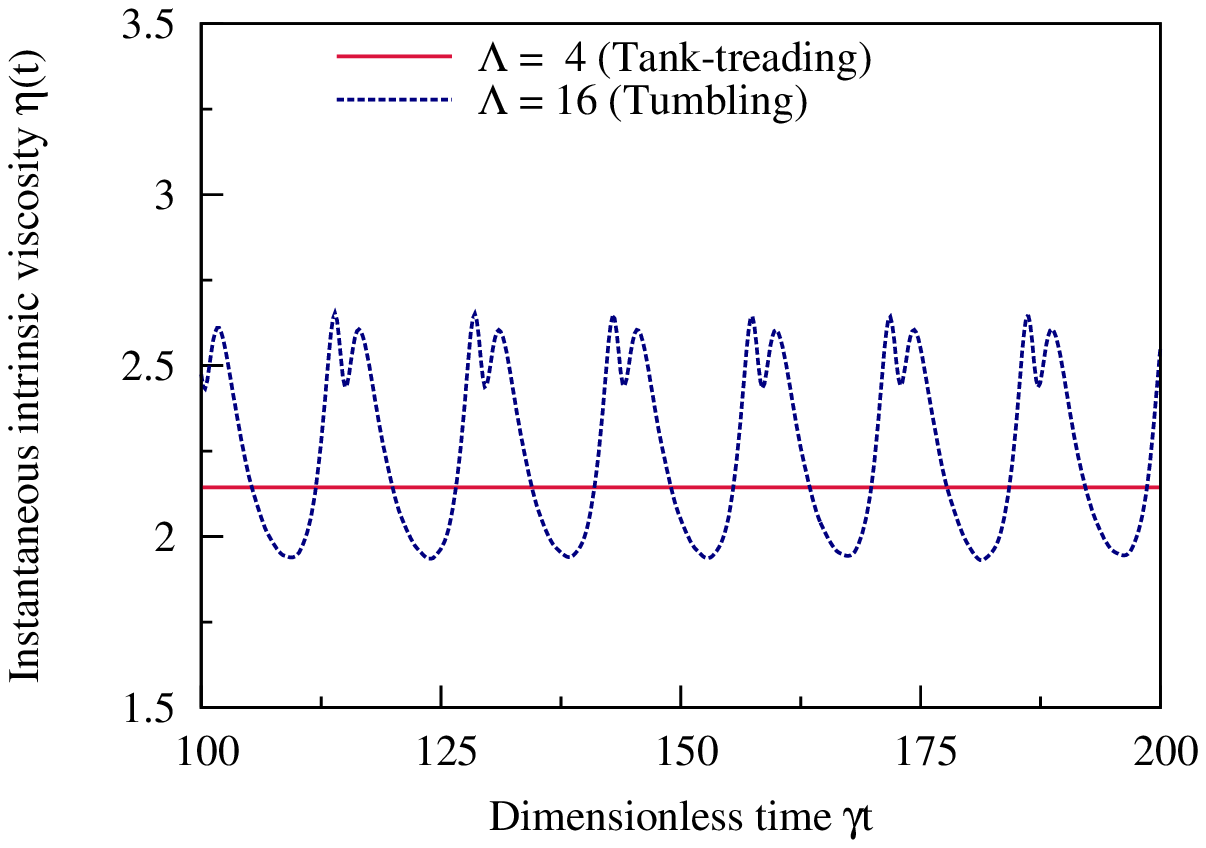}}
\caption{\label{fig:Figure01} (a) Evolution in time (rescaled time $\gamma t$)
of the inclination angle $\theta$ (in degrees) of two vesicles: one in the TT
state ($\Lambda = 4$) and the other one in the TB state ($\Lambda = 16$). The
tank-treading vesicle assumes a steady inclination angle while the tumbling
vesicle assumes a periodic angle. (b) The corresponding evolution in time of the
instantaneous intrinsic viscosity $\eta (t)$ in the TT state ($\Lambda = 4$) and in the TB
state ($\Lambda = 16$). The vesicle in the TT regime performs steady motion and
so does its viscosity. For the TB vesicle, the viscosity evolves in a periodic
manner in time. The two figures recover qualitatively the same behavior as reported
in Ref.~\cite{Ghigliotti2010} computed with 2D boundary integral method
simulations, in the limit of an unbounded suspending fluid ($\chi = 0$). Here, the
confinement is set to $\chi = 0.2$. Other parameters are: ${\rm Re}=0.5$, ${\rm
Ca} = 10$, $\Delta = 0.8$}
\end{figure*}
\textbf{\textit{Parameters}} --- We consider a simulation box with size
$L_x=L_y=200 (\textit{lattice units})$ that represents the suspending fluid and boundary conditions
as introduced above. 
In all simulations, we generate a linear shear flow with a fixed shear rate $\gamma = 2.083
\times 10^{-4}(lu)$. $N$ vesicles which are characterized by
their effective radius $R_0 = 20(lu)$ and their swelling degree $\Delta = 4\pi
A/P^2$ (in 2D) are placed inside the box. The swelling degree is kept at
$\Delta = 0.8$ if not stated otherwise. The conservation of $P$ and $A$ in our
numerical scheme is achieved using: (i) the Lagrangian multiplier field in
Eq.~\ref{eq:membrane}: $\zeta (s,t) = \kappa _P \left[\Delta s(s,t) - \Delta
s(s,t_0)\right]$, where $\Delta s(s,t)$ and $\Delta s(s,t_0)$ are the distance
between two adjacent membrane nodes at time $t$ and $t_0$, respectively (ii)
even though the enclosed fluid is to good approximation an incompressible Newtonian fluid, slight variations of $A$ are observed because of numerical errors \cite{Kaoui2011}.  To further ensure the conservation of $A$ an additional term $\kappa _A \left(A - A_0\right) {\bf n}$ is introduced in Eq.~\ref{eq:membrane}, where $A_0$ is the initial area of a vesicle.
We set $\kappa _P=3(lu)$ and $\kappa _A=0.01(lu)$.\\

Below we present our results as a function of six dimensionless control parameters:
\begin{enumerate}
\item \textit{The Reynolds number} ${\rm Re} = \rho _0  \gamma R_0^2 / \eta _0$, which quantifies the importance of the inertial forces versus the viscous forces; $\rho _0$ is the density of the suspending fluid. We keep ${\rm Re} = 0.5$ in all simulations.
\item \textit{The capillary number} ${\rm Ca} = \eta _0 \gamma R_0^3 / \kappa$, which gives
the ratio between viscous and bending forces. ${\rm Ca}$ is a measure for the deformability of a vesicle. Larger ${\rm Ca}$ leads to larger deformations.
\item \textit{The viscosity contrast} $\Lambda$.
\item \textit{The concentration} $\phi$ of the vesicles in a suspension  $\phi = NA/L_xL_y$.
\item \textit{The swelling degree} $\Delta$.
\item \textit{The degree of confinement} $\chi = 2R_0/L_y = 0.2$. For this
degree of confinement, the tank-treading-to-tumbling transition is expected to
take place at a value of $\Lambda _C = 7.8$~\cite{Kaoui2012}.
\end{enumerate} 
\section{Results} 
\subsection{Rheology of a fluid containing a single vesicle}
We validated our computational method against the case of a single isolated
vesicle ($N=1$) which corresponds to the limit of a very dilute suspension
($\phi \rightarrow 0$). This case was previously studied numerically by
Ghigliotti \textit{et al.}~\cite{Ghigliotti2009,Ghigliotti2010} in 2D and in the
limit of unbounded flow ($\chi = 0$). 

Here, we place a vesicle with $\Delta = 0.8$ in a channel with confinement
$\chi = 0.2$. We use the inclination angle $\theta$ -- the angle between the
flow direction and the main long axis of the vesicle -- as an order parameter
to classify if a vesicle is in the TT ($\theta$ is steady in time) or in the TB
state ($\theta$ varies periodically in time). Fig.~\ref{fig:Figure01a} shows
the evolution in time ($\gamma t$) of the inclination angle $\theta$ of a
vesicle in the TT state ($\Lambda = 4$) and another one in the TB state
($\Lambda = 16$). We compute the instantaneous intrinsic viscosity $\eta(t)$ of the fluid
suspending each of these two vesicles using Eq.~\ref{eq:etat} and we show
in Fig.~\ref{fig:Figure01b} how $\eta(t)$ evolves in time. The vesicle in the TT
state performs a steady motion, therefore, the hydrodynamic stresses exerted on
the bounding walls remain also steady in time. This is why $\eta(t)$ does not
change in time for a tank-treading vesicle. For the tumbling vesicle, the
hydrodynamic stresses on the walls vary periodically in time with the same
frequency as the inclination angle $\theta$. This is the reason for the
viscosity exhibiting a time-dependent behavior for a tumbling vesicle. The
rheological behavior and its correlation with the dynamics reported in
Fig.~\ref{fig:Figure01} are consistent with the results of
Ref.~\cite{Ghigliotti2010}.
 
\begin{figure}
\resizebox{\columnwidth}{!}{\includegraphics{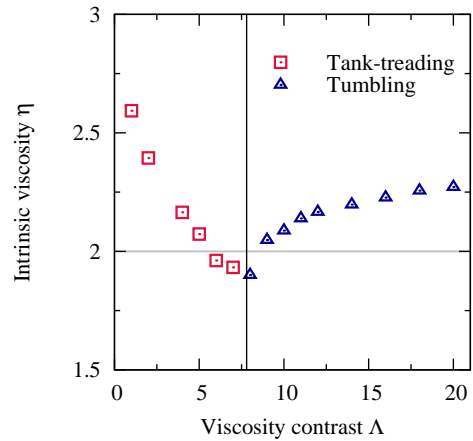}}
\caption{\label{fig:Figure02}
The intrinsic viscosity $\eta$ of a fluid suspending a single vesicle versus
the viscosity contrast $\Lambda$. As $\Lambda$ increases in the TT regime
$\eta$ decreases until it reaches a minimum at a value of $\Lambda$ that
corresponds to the point of the transition from TT to TB. Beyond this critical
value, $\eta$ increases with $\Lambda$ in the TB regime. This non-monotonic
behavior of $\eta$ towards increasing $\Lambda$ agrees with the analytical and
numerical works performed in the unbounded limit ($\chi =0$) and for a single
vesicle (dilute limit $\phi \rightarrow 0$)~\cite{Danker2007,Ghigliotti2010}.
The horizontal line at $\eta = 2$ is the Einstein coefficient in
2D~\cite{Einstein1906,Ghigliotti2010}. Other parameters: ${\rm Re}=0.5$, ${\rm
Ca}=10$, $\Delta = 0.8$ and $\chi = 0.2$.
}
\end{figure}

Using Eq.~\ref{eq:etaI} we obtain the average intrinsic viscosity $\eta$ which is
reported as a function of $\Lambda$ in Fig.~\ref{fig:Figure02}. The horizontal
line at $\eta = 2$ is the Einstein coefficient~\cite{Einstein1906} in
2D~\cite{Ghigliotti2010} (the intrinsic viscosity of an unbounded fluid
suspending a single rigid spherical particle). We see that $\eta$ behaves
differently depending on whether the vesicle performs TT or TB. $\eta$
decreases with $\Lambda$ in the TT regime, because $\theta$ decreases with
$\Lambda$. The vesicle aligns with the flow direction and thus its surrounding
fluid experiences less resistance. $\eta$ continues to decrease with $\Lambda$
until it drops down to a minimum at exactly $\Lambda _{C}$, the critical
viscosity contrast at which the TT-TB transition occurs. Beyond this critical
point, in the tumbling regime, $\eta$ increases again with $\Lambda$. For a
tumbling vesicle, the viscous dissipation and the stresses on the walls
increase with $\Lambda$. This leads to the increase of $\eta$ with $\Lambda$.
Fig.~\ref{fig:Figure02} clearly shows that the intrinsic viscosity $\eta$ of a
fluid containing a single vesicle is a non-monotonic function of the viscosity
contrast $\Lambda$.  It is a decreasing function in the TT regime and an
increasing function in the TB regime. It changes its behavior at a minimum that
coincides with the critical transition point of the TT-TB transition. This
observation agrees (qualitatively) with the previous analytical work of Danker
and Misbah~\cite{Danker2007} and the numerical work of Ghigliotti \textit{et
al.}~\cite{Ghigliotti2010}. In contrast to the work of those authors our
Reynolds number ${\rm Re} = 0.5$ is not zero and the vesicle is confined
(${\chi} = 0.2$), but we also recover the predicted non-monotonic behavior of
$\eta$ versus $\Lambda$.
\subsection{Effect of concentration $\phi$}
The concentration $\phi$ of a suspension is varied by increasing/decreasing the
number of its vesicles $N$, while keeping the size of the simulation box
constant. We consider three suspensions with concentrations $\phi = 7.5\%,
15.1\%$ and $22.6\%$ corresponding to $3$, $6$ and $9$ vesicles, respectively. 
\begin{figure}
\resizebox{\columnwidth}{!}{\includegraphics{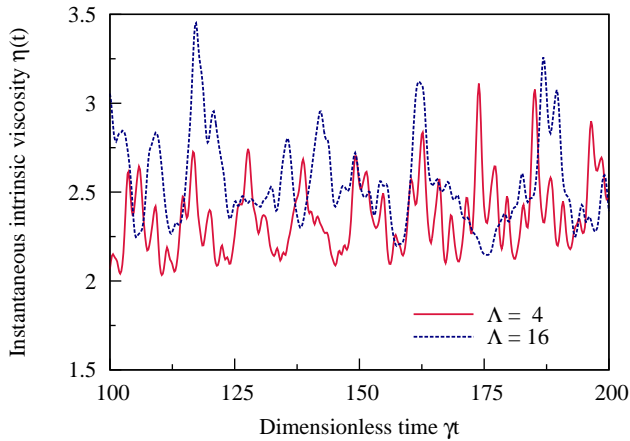}}
\caption{\label{fig:Figure03}
Evolution in time of the instantaneous intrinsic viscosity $\eta (t)$ of two suspensions with
the same concentration $\phi = 15.1\%$ (that corresponds to $6$ vesicles), but
with different viscosity contrast: $\Lambda = 4$ and $\Lambda = 16$. $\eta(t)$
fluctuates because of the motion and the ordering of the vesicles in
response to the applied external shear flow. The large peaks in $\eta(t)$ are
caused by the events of the vesicle-vesicle and vesicle-wall hydrodynamic
collisions. Other parameters: ${\rm Re=0.5}$, ${\rm Ca}=10$, $\Delta = 0.8$
and $\chi = 0.2$.
}
\end{figure}
After the system has reached a quasi steady state, we measure the intrinsic
viscosity $\eta(t)$. Fig.~\ref{fig:Figure03} depicts how $\eta(t)$ evolves in time
for two suspensions having the same $\phi = 15.1\%$, but different viscosity
contrast: $\Lambda = 4$ and $16$. For both cases, $\eta(t)$ evolves in an
unsteady way, even for the non-tumbling vesicles (for $\Lambda = 4$). It
largely fluctuates and sometimes shows higher sharp peaks. These fluctuations
are correlated with how the vesicles rearrange themselves in response to the
applied flow.
\begin{figure*}
\centering
\subfloat[$\gamma t = 179.14$]{\label{fig:Figure04a}\includegraphics[angle=-90,width=0.15\textwidth]{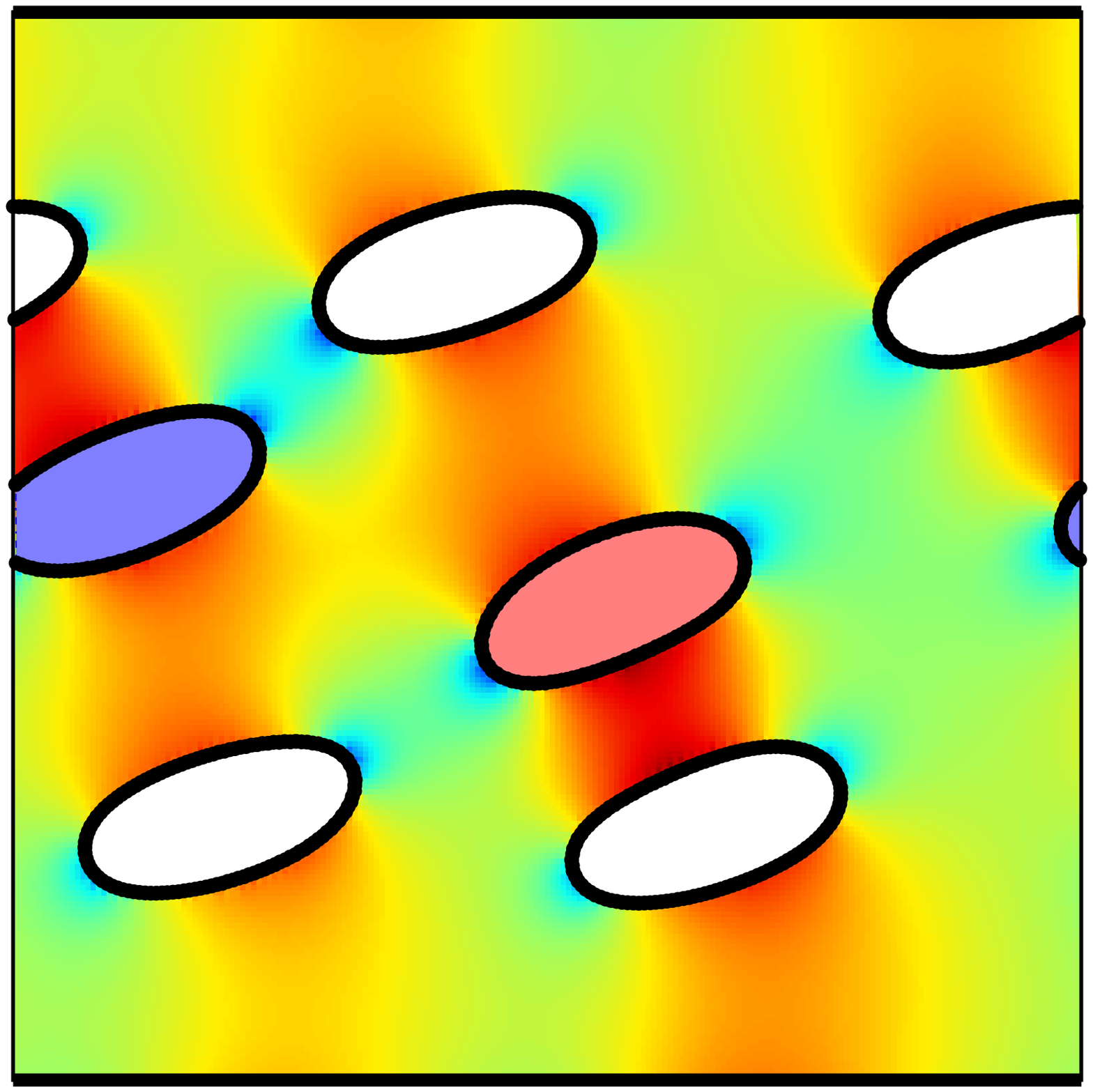}}
%t = 179.138
\quad
\subfloat[$\gamma t = 181.22 $]{\label{fig:Figure04b}\includegraphics[angle=-90,width=0.15\textwidth]{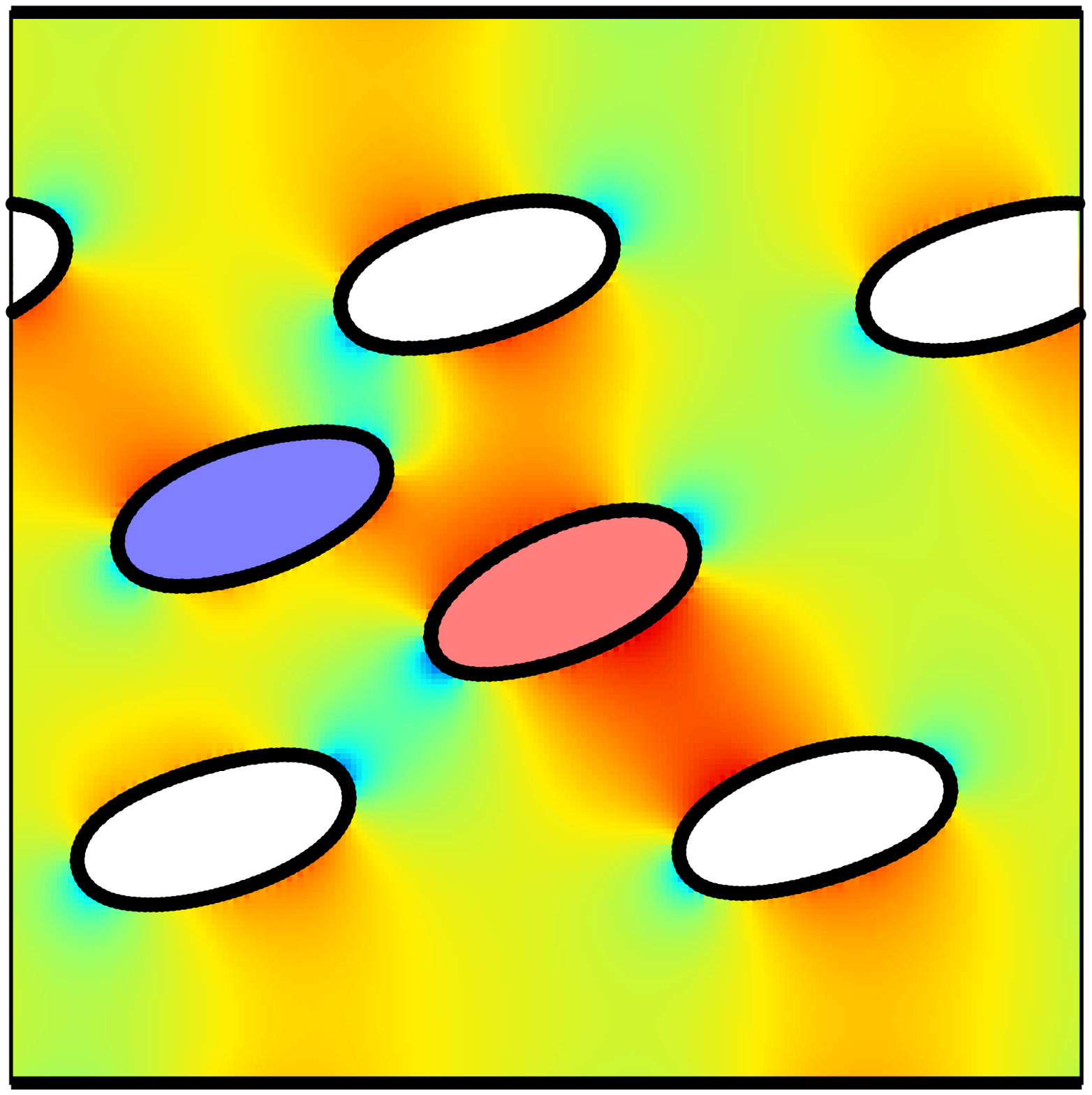}}
%t = 181.221
\quad
\subfloat[$\gamma t = 183.30 $]{\label{fig:Figure04c}\includegraphics[angle=-90,width=0.15\textwidth]{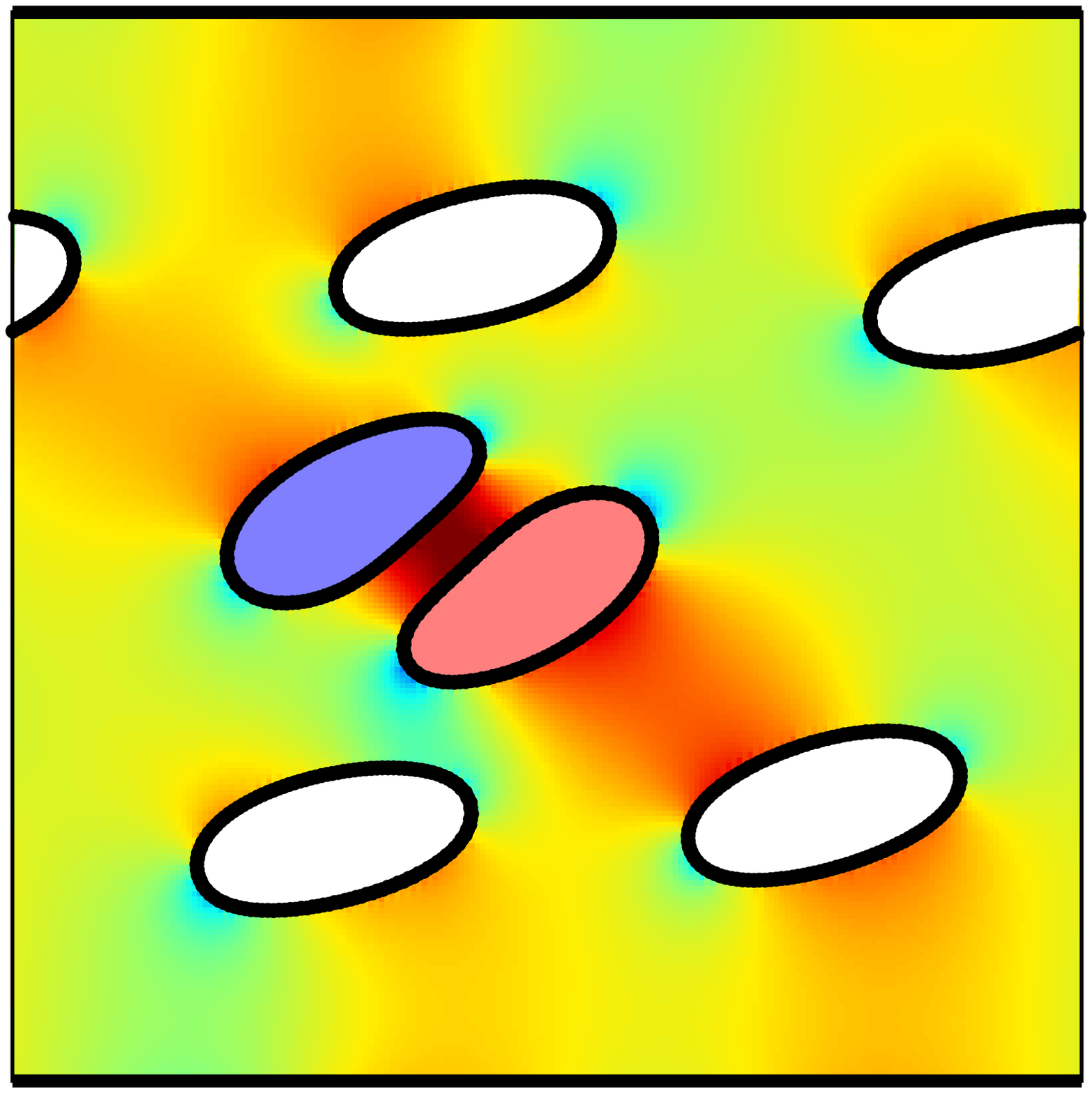}}
%t = 183.304
\quad
\subfloat[$\gamma t = 185.39$]{\label{fig:Figure04d}\includegraphics[angle=-90,width=0.15\textwidth]{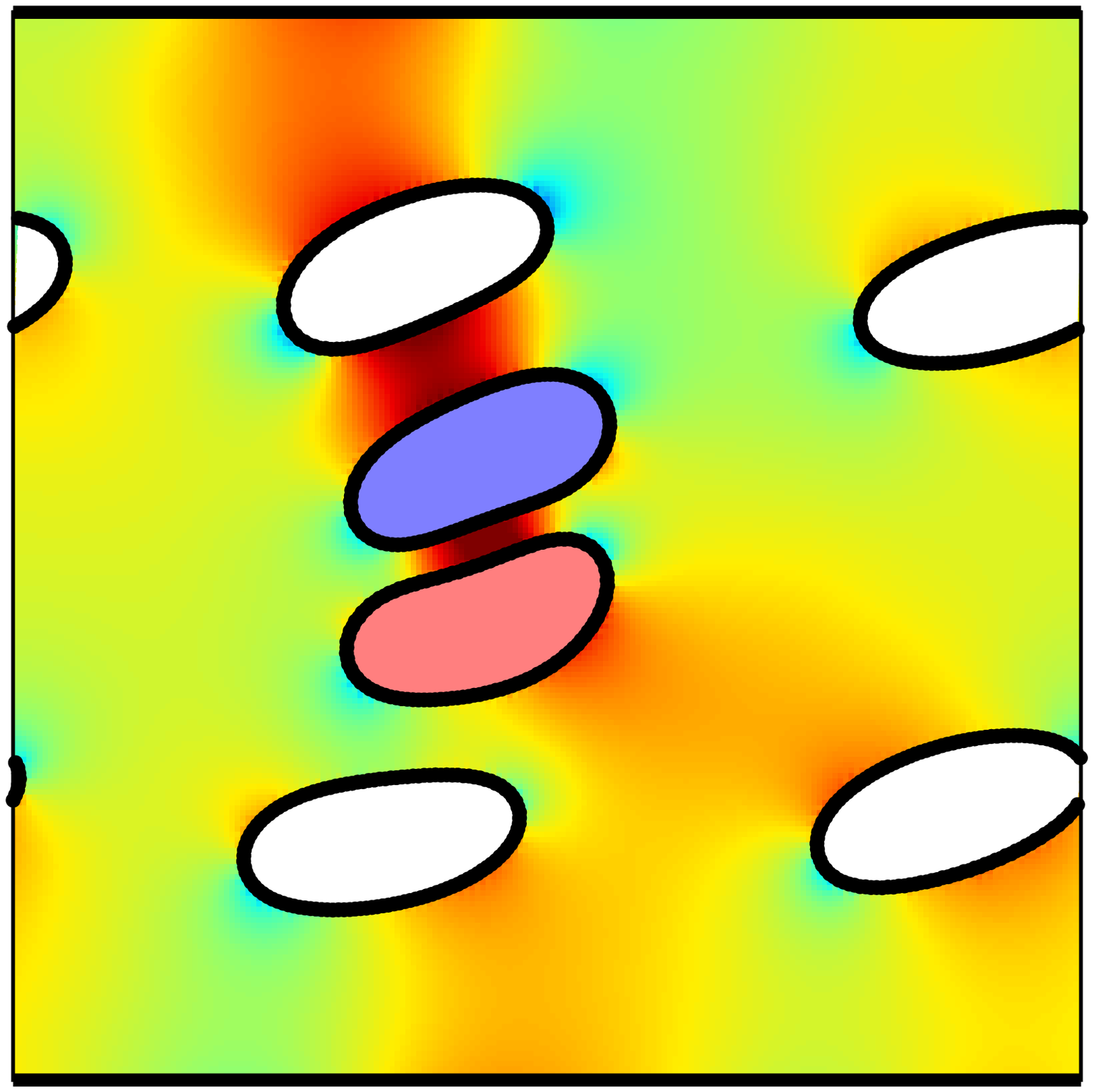}}
%t = 185.387
\quad
\subfloat[$\gamma t = 187.47$]{\label{fig:Figure04e}\includegraphics[angle=-90,width=0.15\textwidth]{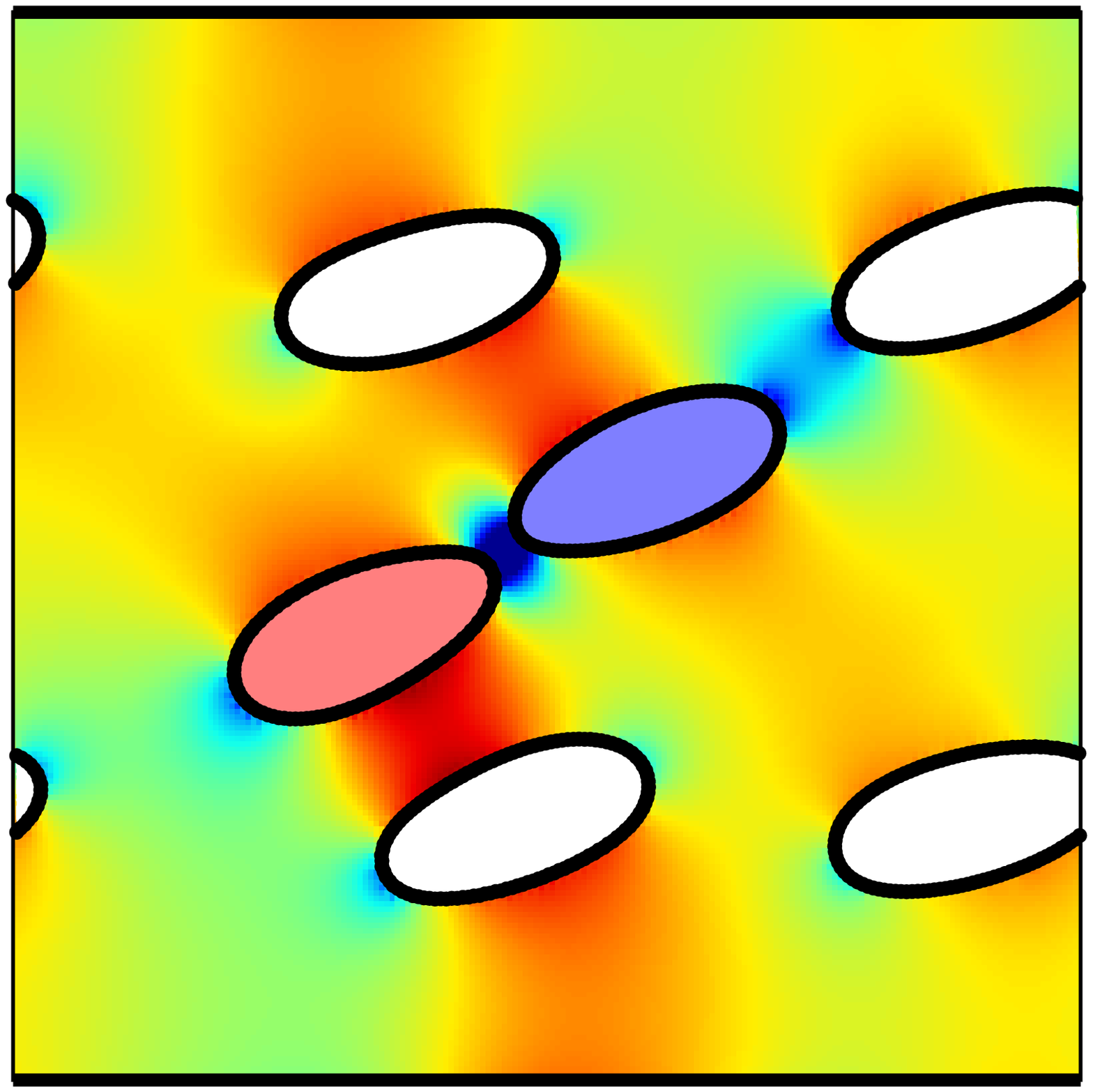}}
%t = 187.470
\quad
\subfloat[$\gamma t = 189.55$]{\label{fig:Figure04f}\includegraphics[angle=-90,width=0.15\textwidth]{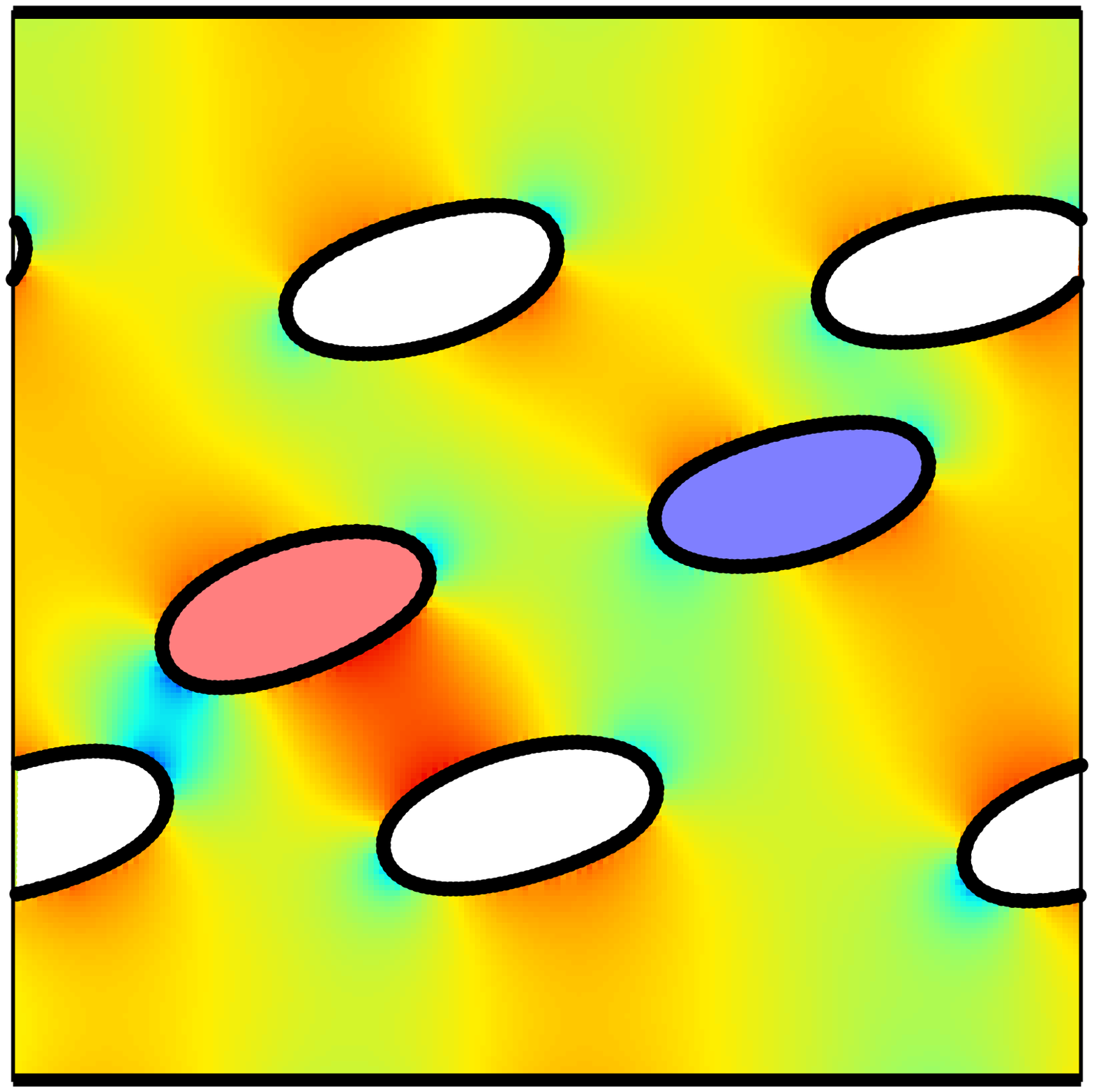}}
%t = 189.553
\caption{\label{fig:Figure04} Snapshots taken at equal time intervals showing
the motion of six vesicles ($\phi = 15.1\%$) with a viscosity contrast
$\Lambda = 4$ in shear flow. The flow direction is from left to right. The
background color map shows the pressure field that develops around the
vesicles. Red-colored regions correspond to regions with higher pressure, while
blue-colored ones corresponds to lower pressure. The two blue- and red-colored
vesicles undergo a hydrodynamic collision. All vesicles perform TT with a
steady prolate shape and assume almost the same steady positive inclination
angle. Only the angles of the colliding vesicles vary and reach a maximum
at the moment of the collision (d). Other parameters: ${\rm Re=0.5}$, ${\rm Ca}=10$, $\Delta = 0.8$ and $\chi = 0.2$.}
\end{figure*}
\begin{figure*}
\centering
\subfloat[$\gamma t = 102.90$]{\label{fig:Figure05a}\includegraphics[angle=-90,width=0.15\textwidth]{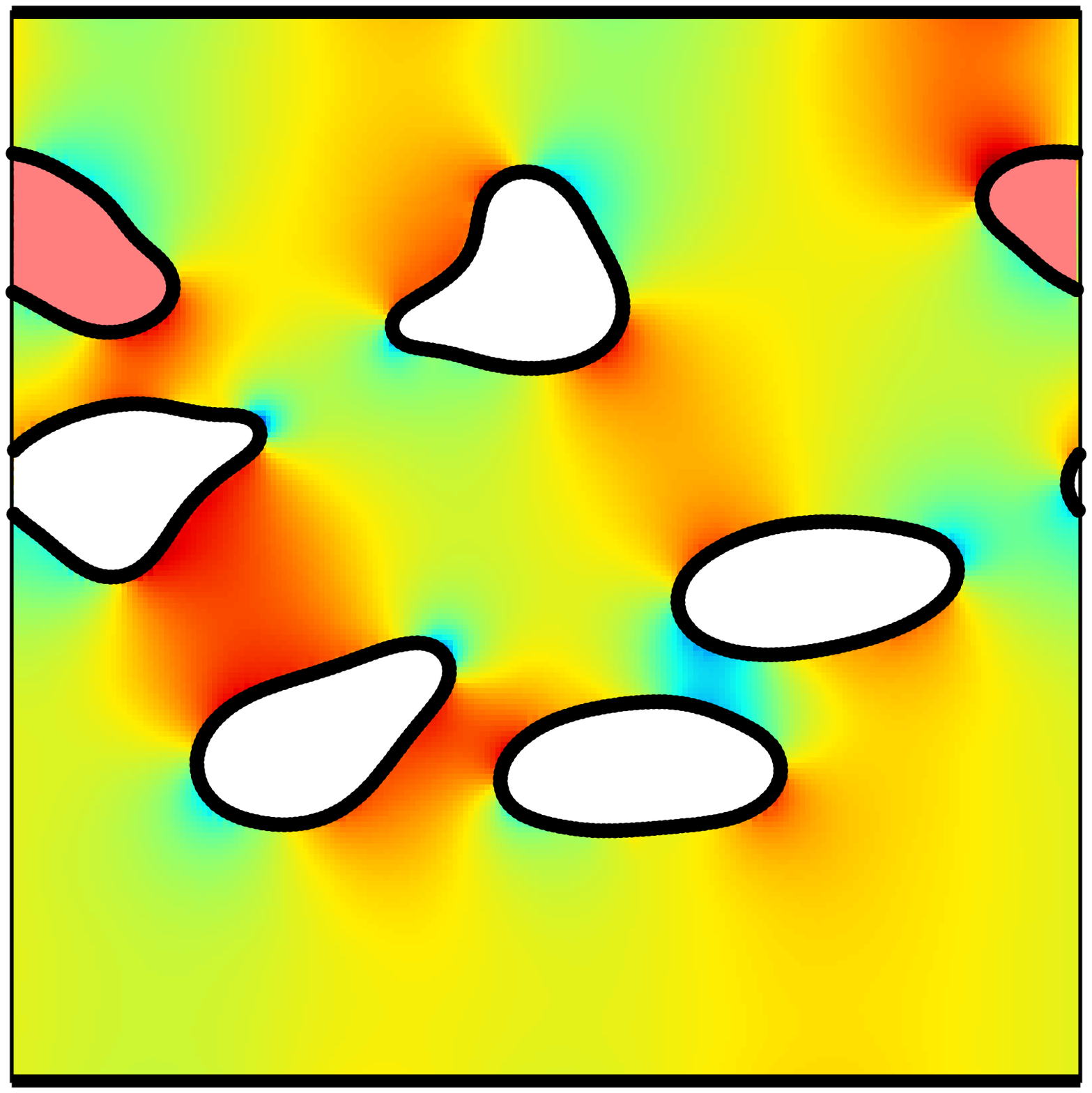}}
%\gamma t = 102.9002
\quad
\subfloat[$\gamma t = 103.32$]{\label{fig:Figure05b}\includegraphics[angle=-90,width=0.15\textwidth]{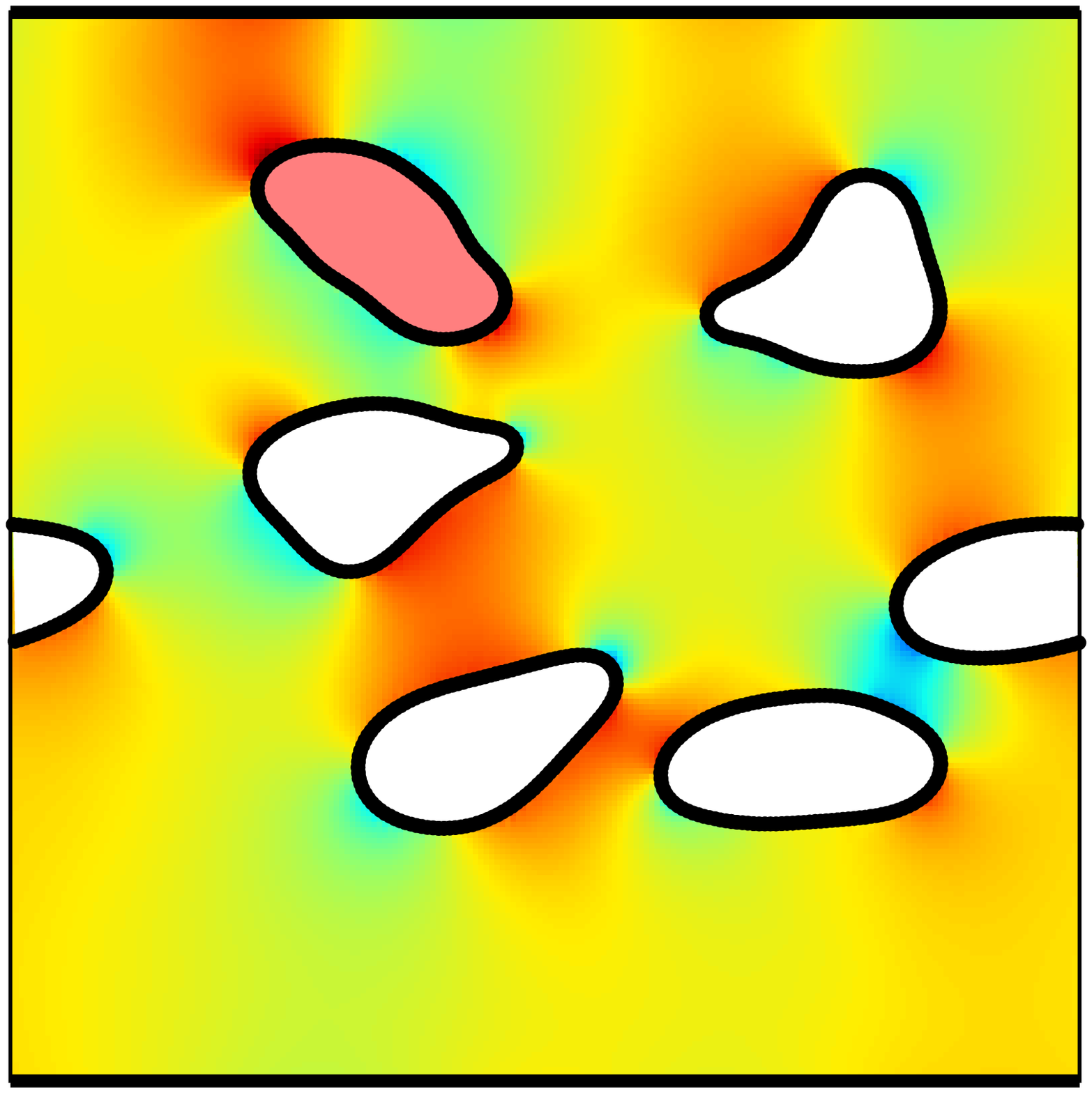}}
%\gamma t = 103.3168
\quad
\subfloat[$\gamma t = 103.73$]{\label{fig:Figure05c}\includegraphics[angle=-90,width=0.15\textwidth]{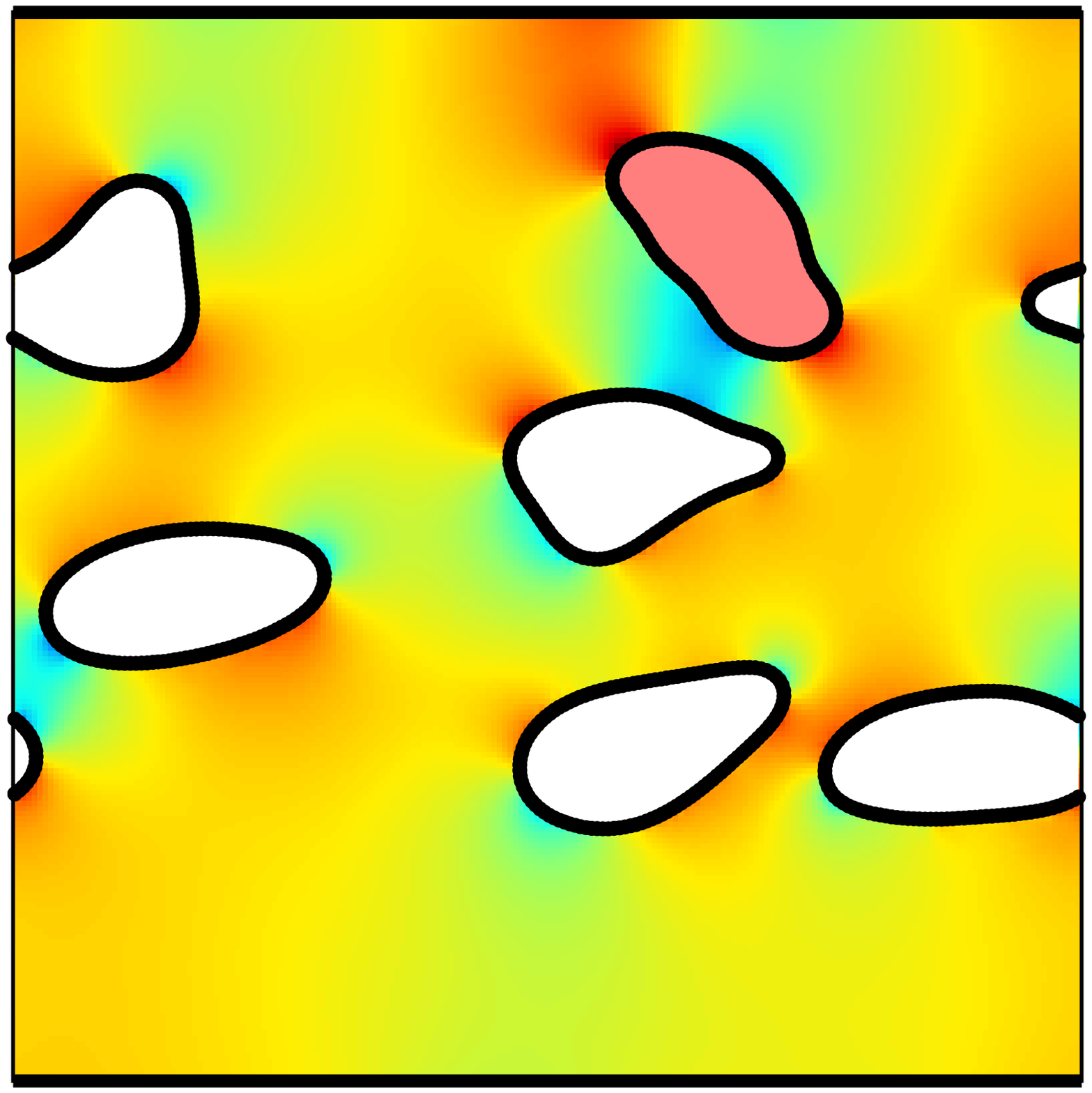}}
%\gamma t = 103.7334
\quad
\subfloat[$\gamma t = 104.15$]{\label{fig:Figure05d}\includegraphics[angle=-90,width=0.15\textwidth]{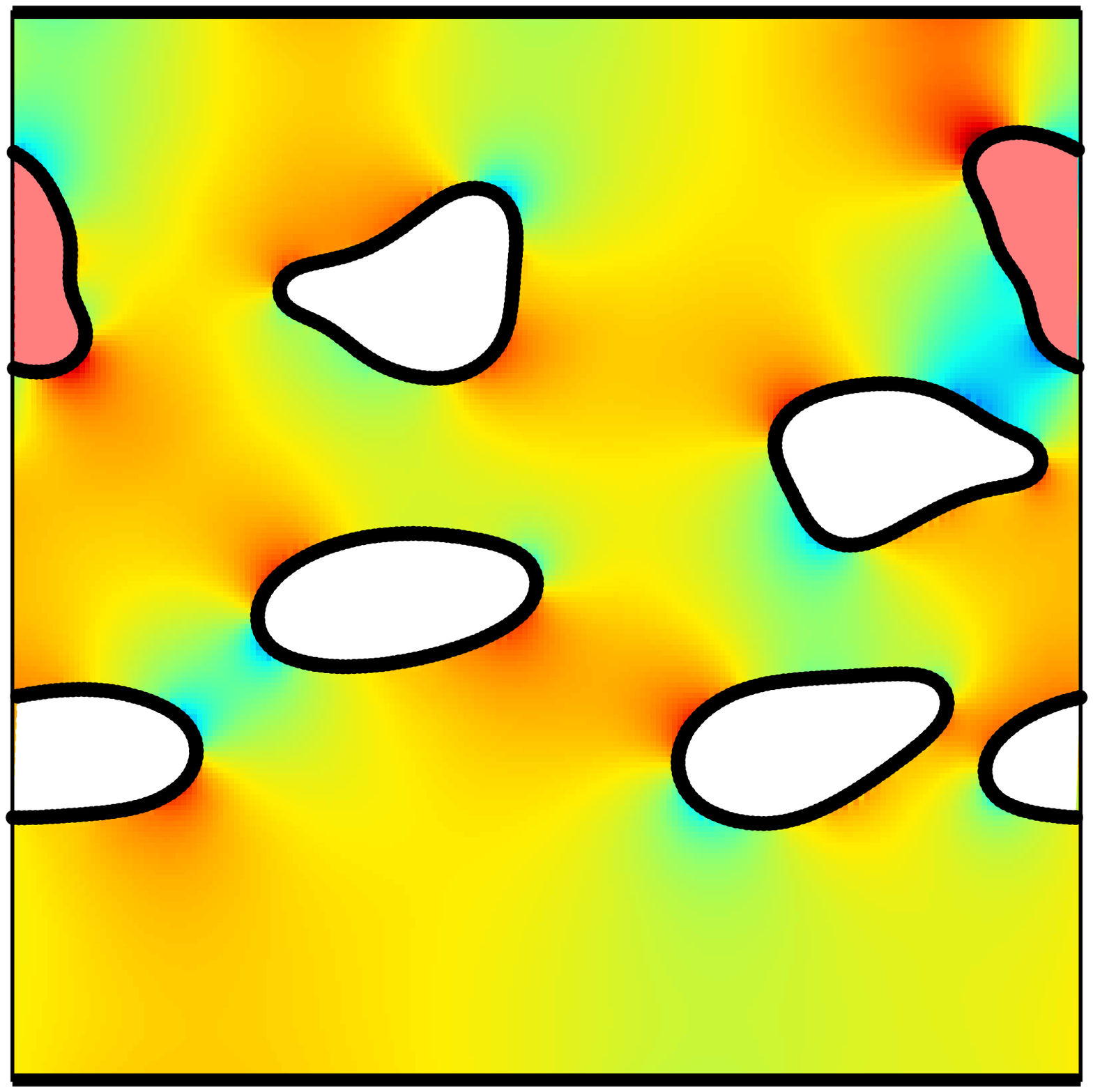}}
%\gamma t = 104.15
\quad
\subfloat[$\gamma t = 104.57$]{\label{fig:Figure05e}\includegraphics[angle=-90,width=0.15\textwidth]{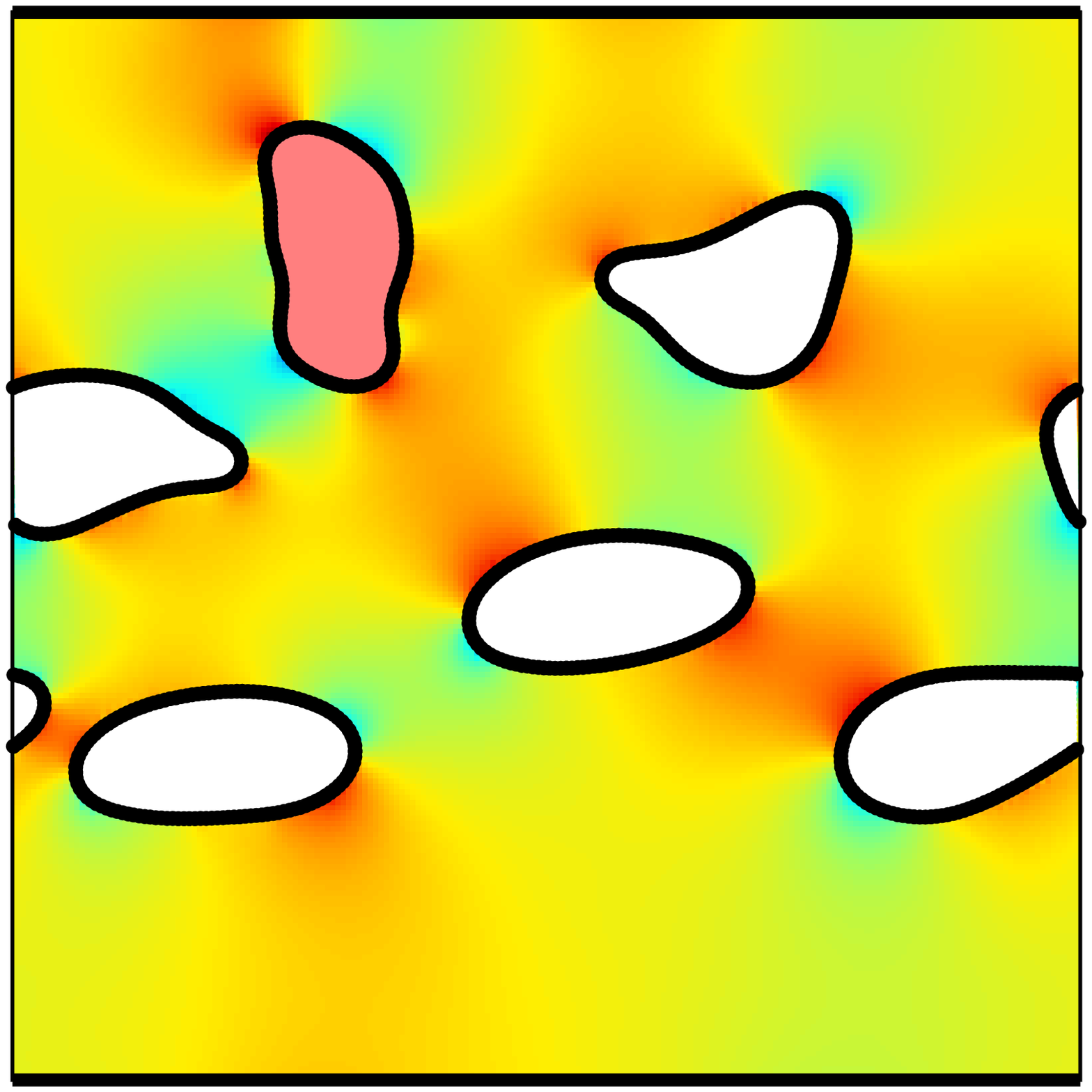}}
%\gamma t = 104.5666
\quad
\subfloat[$\gamma t = 104.98$]{\label{fig:Figure05f}\includegraphics[angle=-90,width=0.15\textwidth]{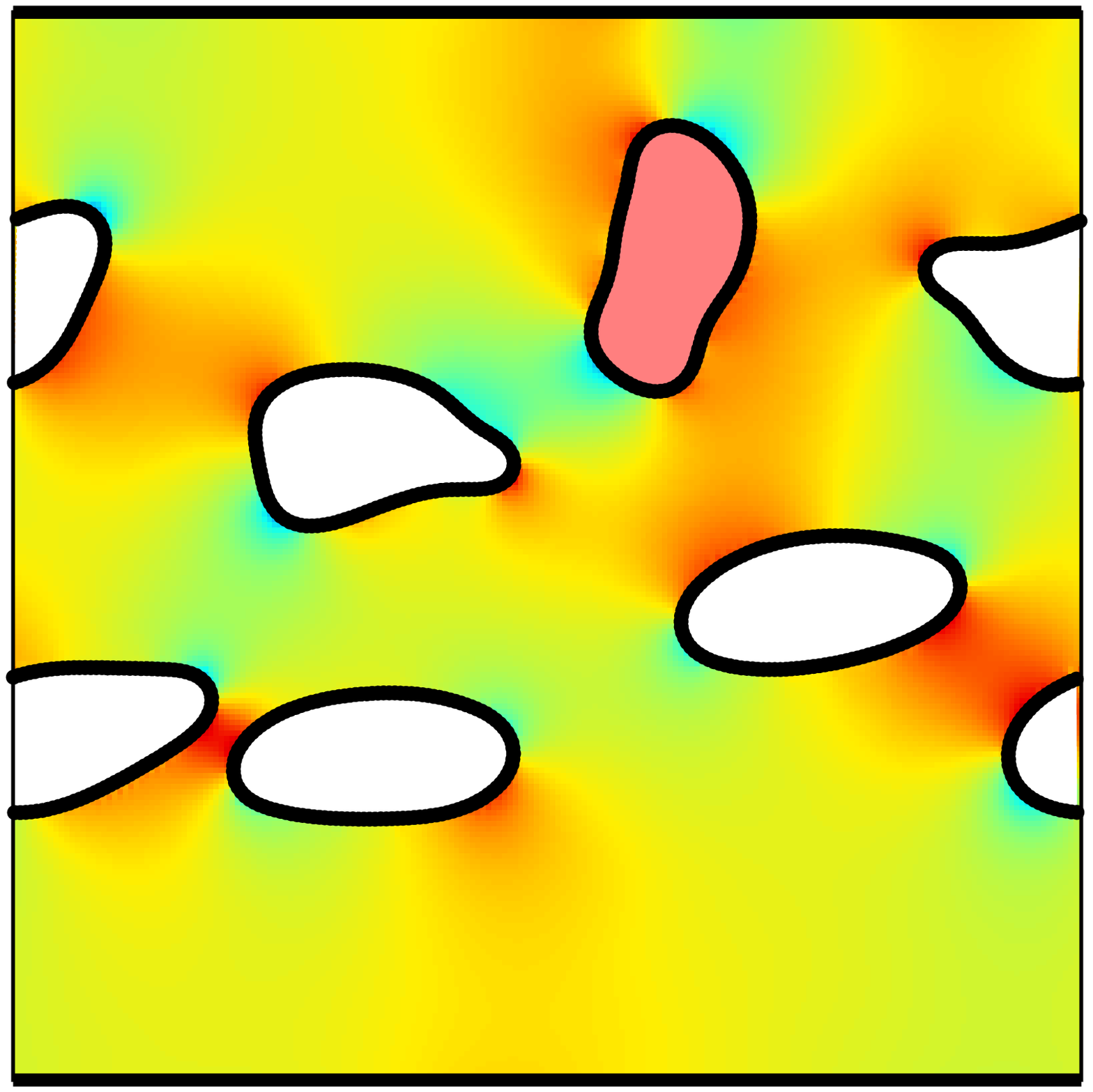}}
%\gamma t = 104.9832
\\
\subfloat[$\gamma t = 105.40$]{\label{fig:Figure05g}\includegraphics[angle=-90,width=0.15\textwidth]{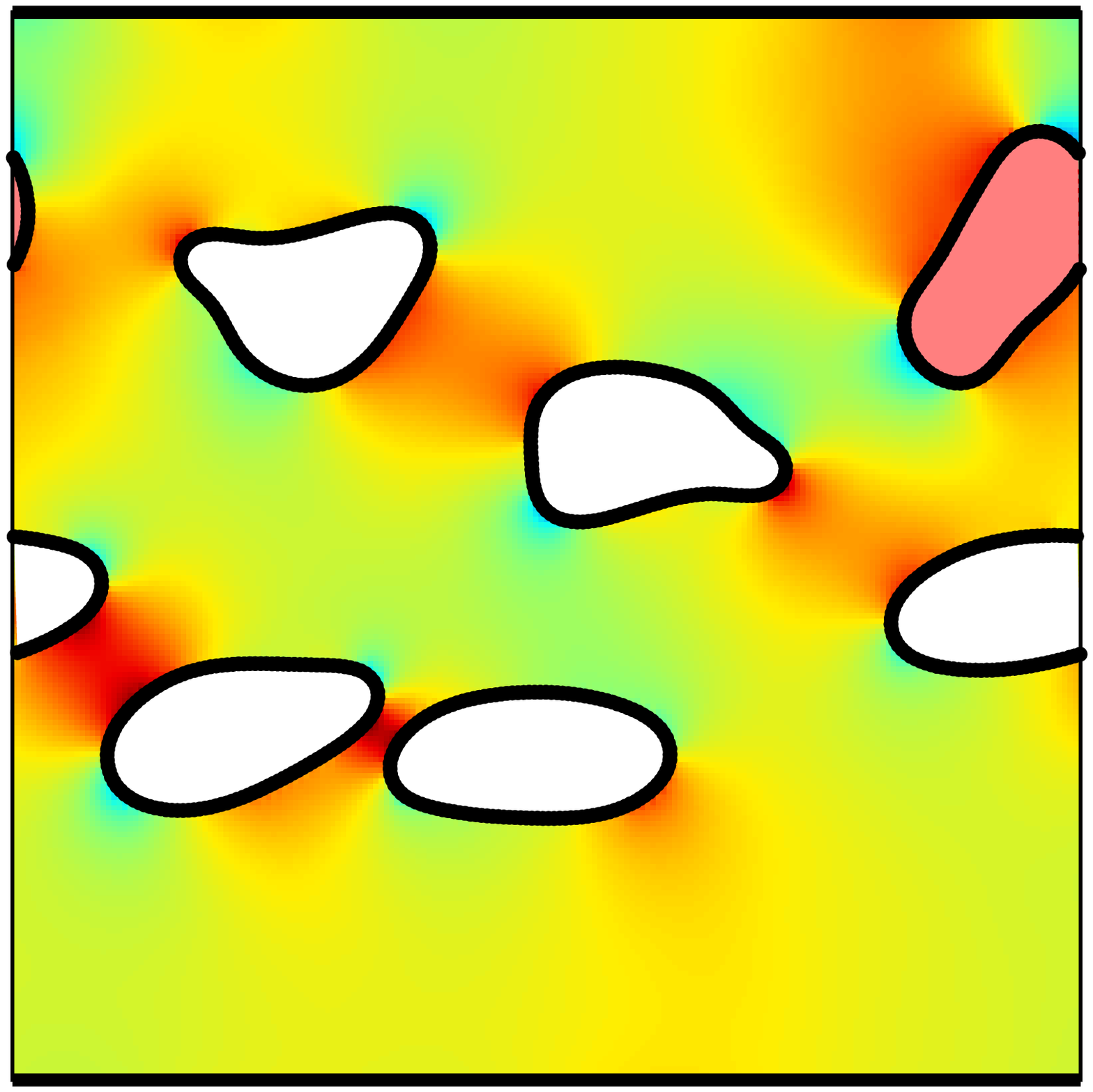}}
%\gamma t = 105.3998
\quad
\subfloat[$\gamma t = 105.82$]{\label{fig:Figure05h}\includegraphics[angle=-90,width=0.15\textwidth]{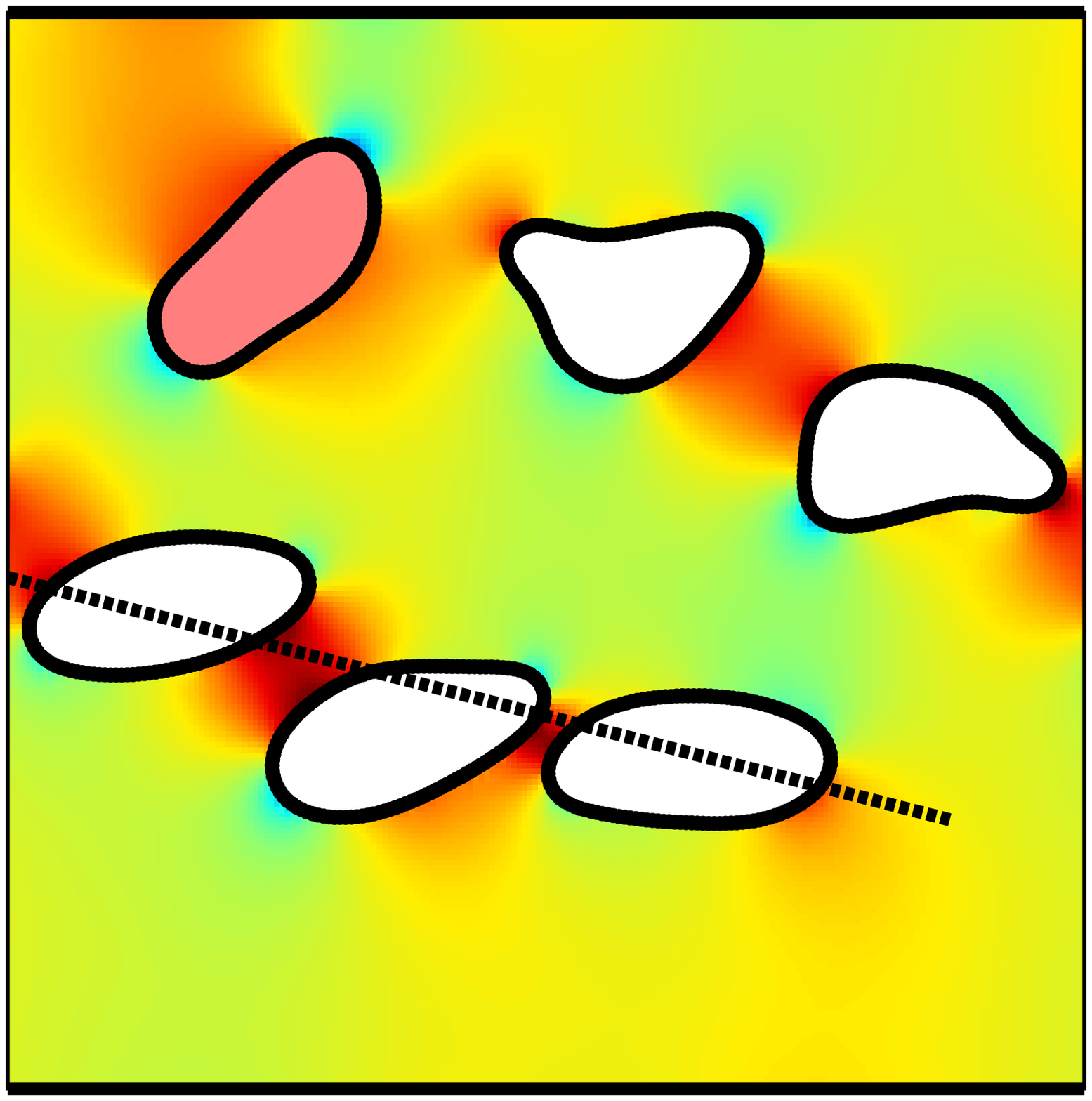}}
%\gamma t = 105.8164
\quad
\subfloat[$\gamma t = 106.23$]{\label{fig:Figure05i}\includegraphics[angle=-90,width=0.15\textwidth]{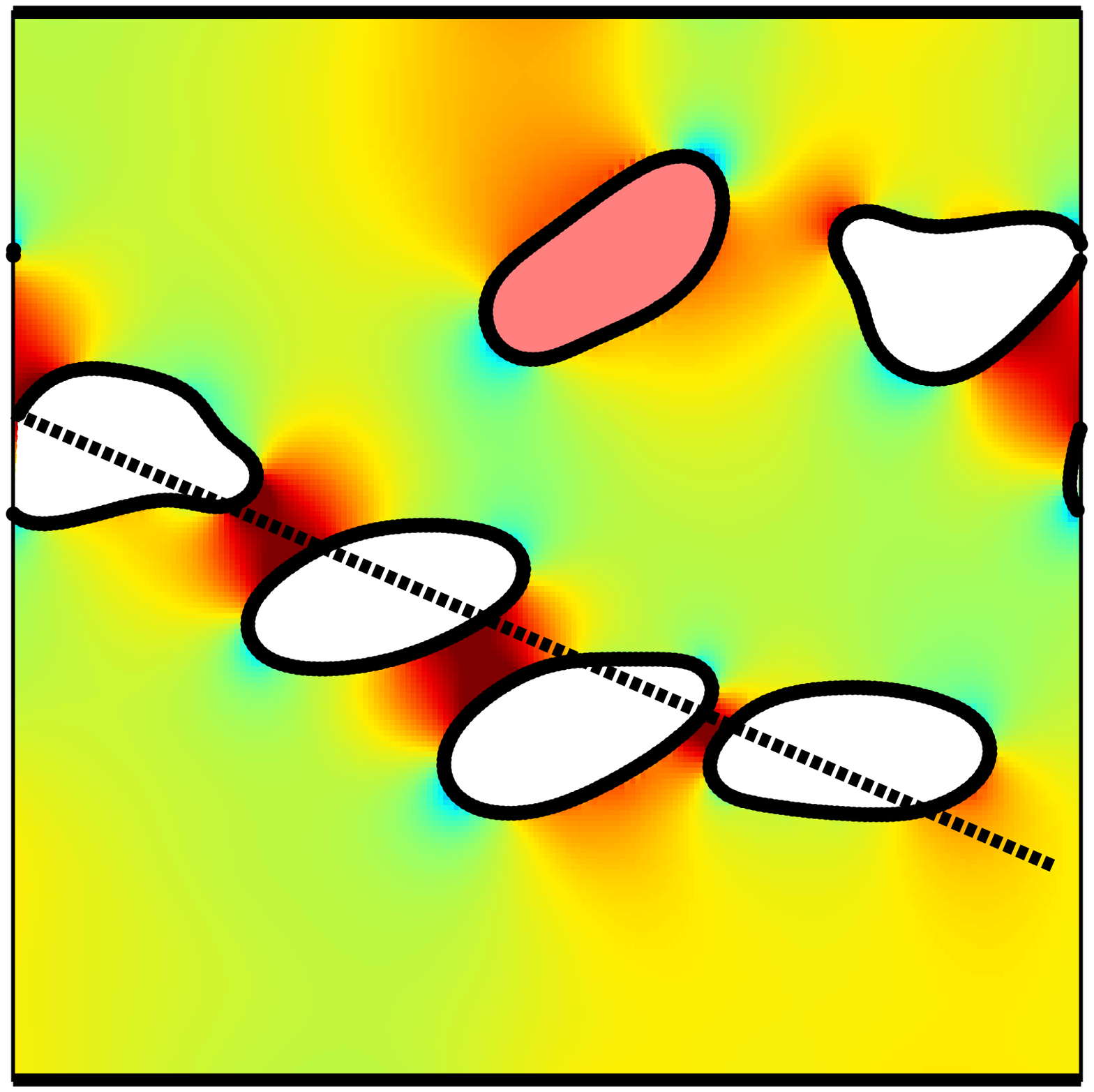}}
%\gamma t = 106.2330
\quad
\subfloat[$\gamma t = 106.65$]{\label{fig:Figure05j}\includegraphics[angle=-90,width=0.15\textwidth]{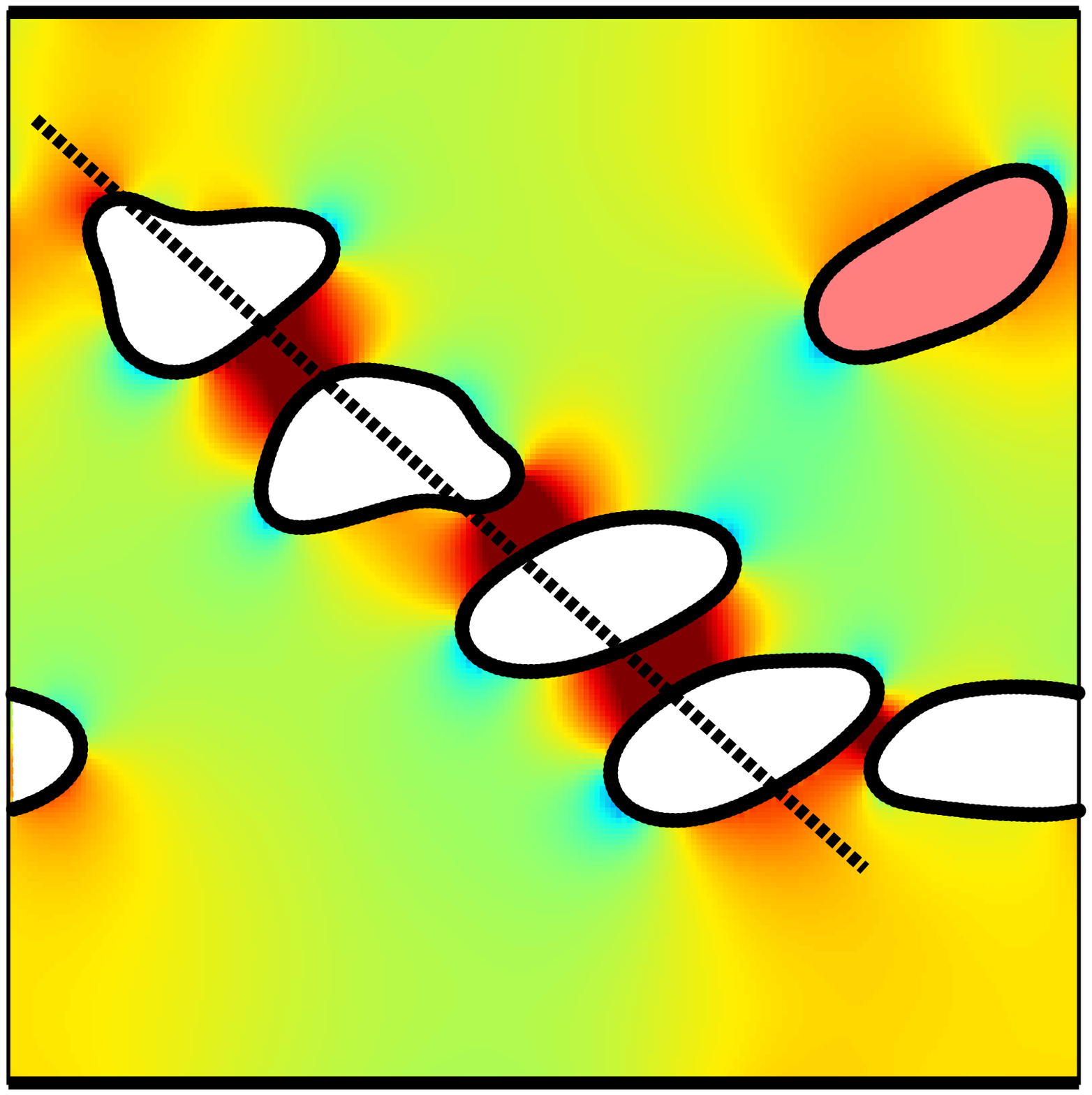}}
%\gamma t = 106.6496
\quad
\subfloat[$\gamma t = 107.07$]{\label{fig:Figure05k}\includegraphics[angle=-90,width=0.15\textwidth]{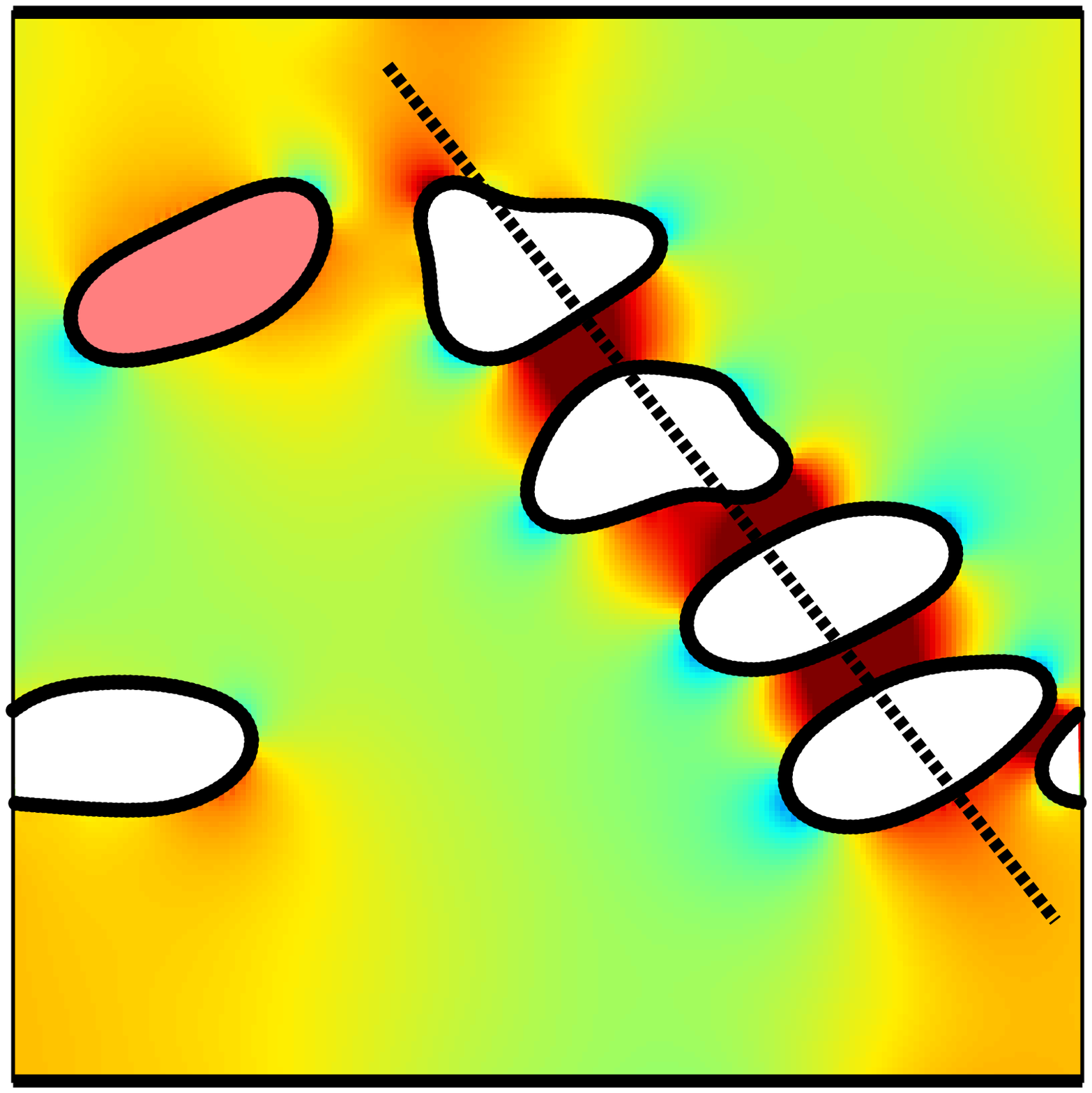}}
%\gamma t = 107.0662
\quad
\subfloat[$\gamma t = 107.48$]{\label{fig:Figure05l}\includegraphics[angle=-90,width=0.15\textwidth]{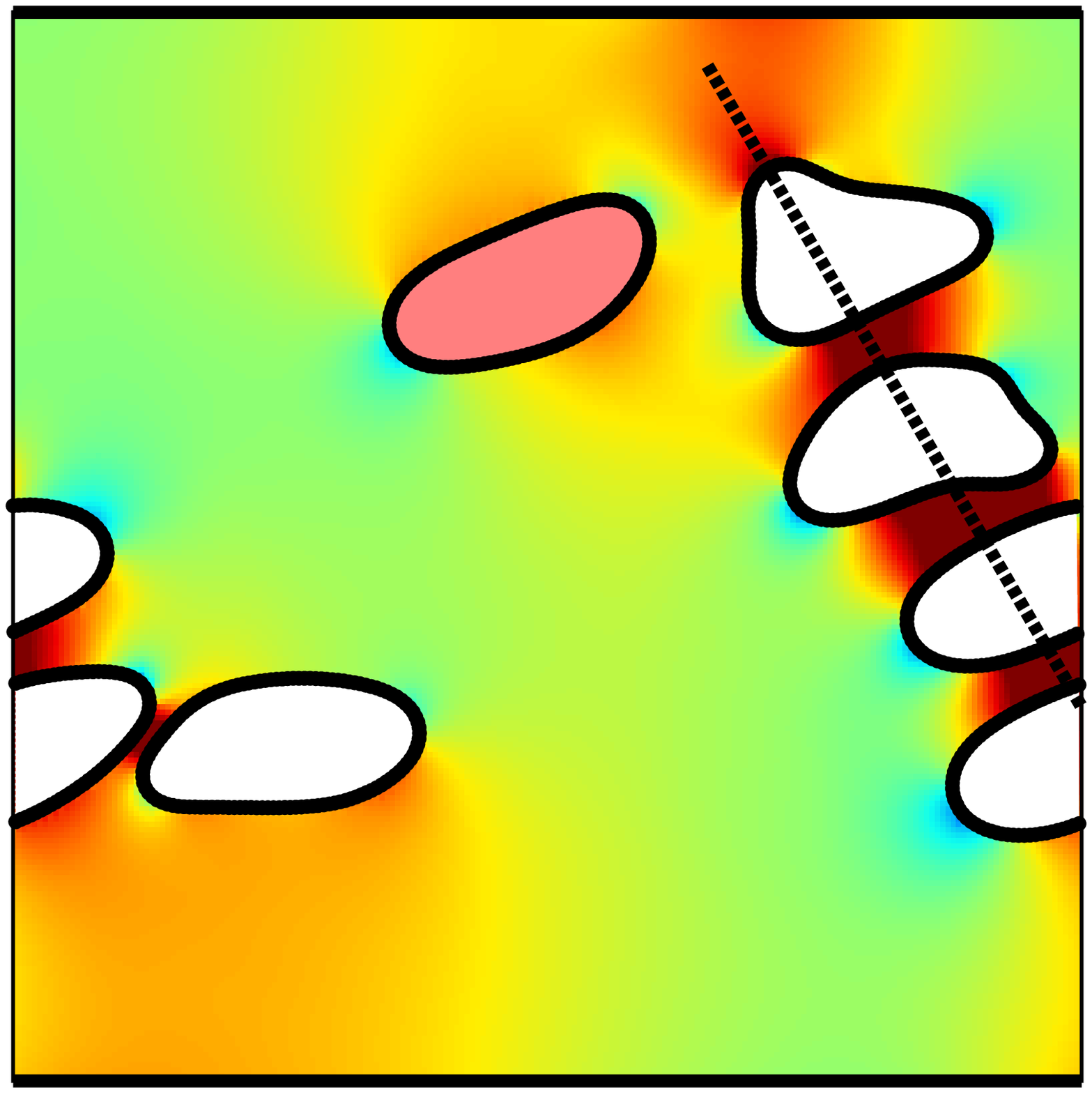}}
%\gamma t = 107.4828
\caption{\label{fig:Figure05} Snapshots taken at equal time intervals showing
the motion of six vesicles ($\phi = 15.1\%$) with viscosity contrast $\Lambda =
16$ in shear flow. The background color map shows the pressure field that
develops around the vesicles. In the time window shown here, all vesicles undergo TT
except the one colored in red. For $\Lambda = 16$, vesicles are expected to
tumble, but because of the confinement and the vesicle-vesicle hydrodynamical
interaction, the tumbling motion is inhibited.  All vesicles adhere to shapes
which largely deviate from the prolate shape a free vesicle would adopt. These
shapes result on the one hand from the collisions that lead to deformations and
on the other hand from the long time they require to recover their equilibrium
shape. Another observed feature is the formation of a rouleau-like structure
of a given number of vesicles that performs a collective tumbling motion. When
the main axis (dashed line) of this chain of vesicles is perpendicular to the
bounding walls it results in an increased effective viscosity.}
\end{figure*}
Fig.~\ref{fig:Figure04} shows snapshots taken at equal time intervals
displaying the motion of vesicles with $\Lambda = 4$. For this low viscosity
contrast, vesicles are in the tank-treading state. They pass each other while
assuming almost the same steady inclination angle. Collisions of hydrodynamic
nature occur occasionally and become more frequent upon increasing
concentration $\phi$. When two vesicles collide, as is the case for the red-
and the blue-colored vesicles in Fig.~\ref{fig:Figure04}, their respective
inclination angle reaches a maximum at the moment of the collision (see
Fig.~\ref{fig:Figure04}c and Fig.~\ref{fig:Figure04}d) as observed also
experimentally~\cite{Kantsler2008,Levant2012}. This event increases the flow resistance,
which is amplified by the presence of the bounding walls. Here, the two
interacting vesicles have no way to move farther away from each other, in
contrast to the case of unbounded suspensions. Thus, they collide later with
other neighboring vesicles or with the walls. This latter effect results in
exerting firm stresses upon the walls and causes the peaks of $\eta(t)$ observed
in Fig.~\ref{fig:Figure03}. In Fig.~\ref{fig:Figure05}, we show snapshots for
the suspension with $\Lambda = 16$, a higher value for which vesicles are
expected to undergo tumbling.  However, we can clearly see that vesicles do not
have sufficient free space around them to tumble. Each vesicle is surrounded
and hindered by others. Thus, vesicles are forced to undergo the tank-treading
motion, although they would tumble in free space. The area available for
vesicles to tumble reduces dramatically when increasing $\phi$. In the time
window represented in Fig.~\ref{fig:Figure05} only one vesicle is tumbling (the red-colored vesicle in the 
top-half of the simulation box). Another observed feature in Fig.~\ref{fig:Figure05} is the 
formation of a rouleau-like structure of a given number of vesicles that performs a collective tumbling 
motion. When the main axis (dashed line) of this chain of vesicles is perpendicular to the bounding walls it 
results in an increased effective viscosity. The formation of the rouleau-like structure is not a usual 
behavior. For example, it has not taken place in the
simulation presented in Fig.~\ref{fig:Figure04}.

\begin{figure}
\resizebox{\columnwidth}{!}{\includegraphics{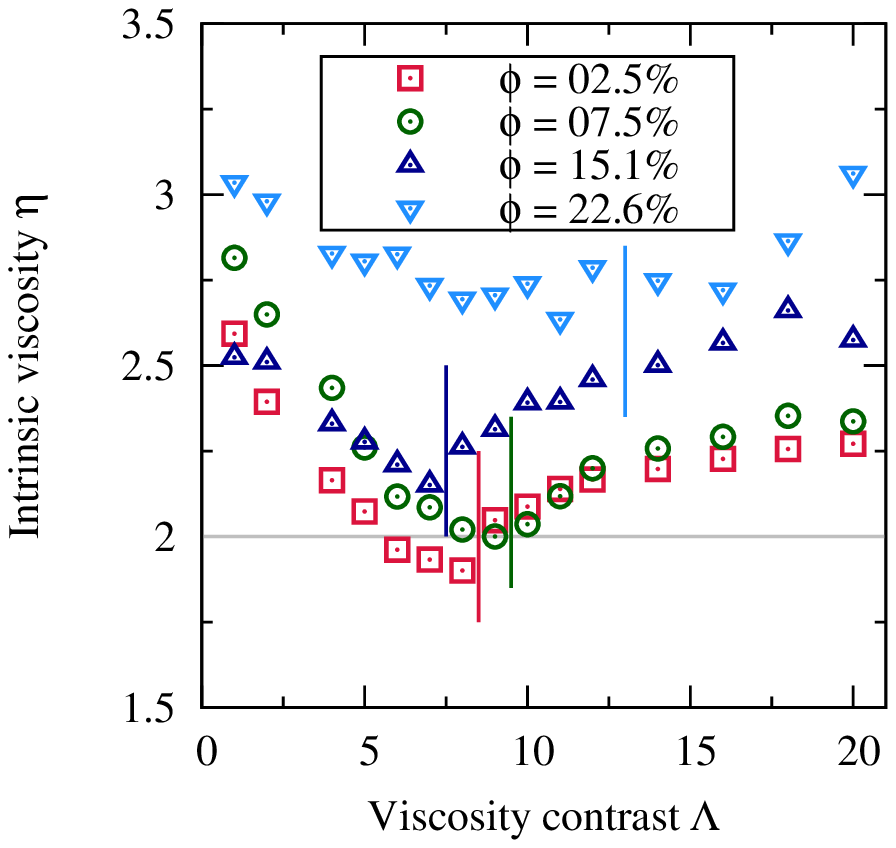}}
\caption{\label{fig:Figure06}
The intrinsic viscosity $\eta$ versus the viscosity contrast $\Lambda$ of four
suspensions with concentrations $\phi = 2.5\%$, $7.5\%$, $15.1\%$
and $22.6\%$. The vertical lines denote the minima of the curves. For every
given value of $\Lambda$, a suspension becomes more and more viscous with
increasing $\phi$. For the dense suspension ($\phi = 22.6\%$), the
non-monotonic behavior of $\eta$ with $\Lambda$ is less pronounced.
Furthermore, the viscosity contrast of the minimum of $\eta$ does not vary in a
monotonic way with $\phi$. Other parameters: ${\rm Re}=0.5$, ${\rm Ca}=10$,
$\Delta = 0.8$ and $\chi = 0.2$.}
\end{figure}
\begin{figure}
\resizebox{\columnwidth}{!}{\includegraphics{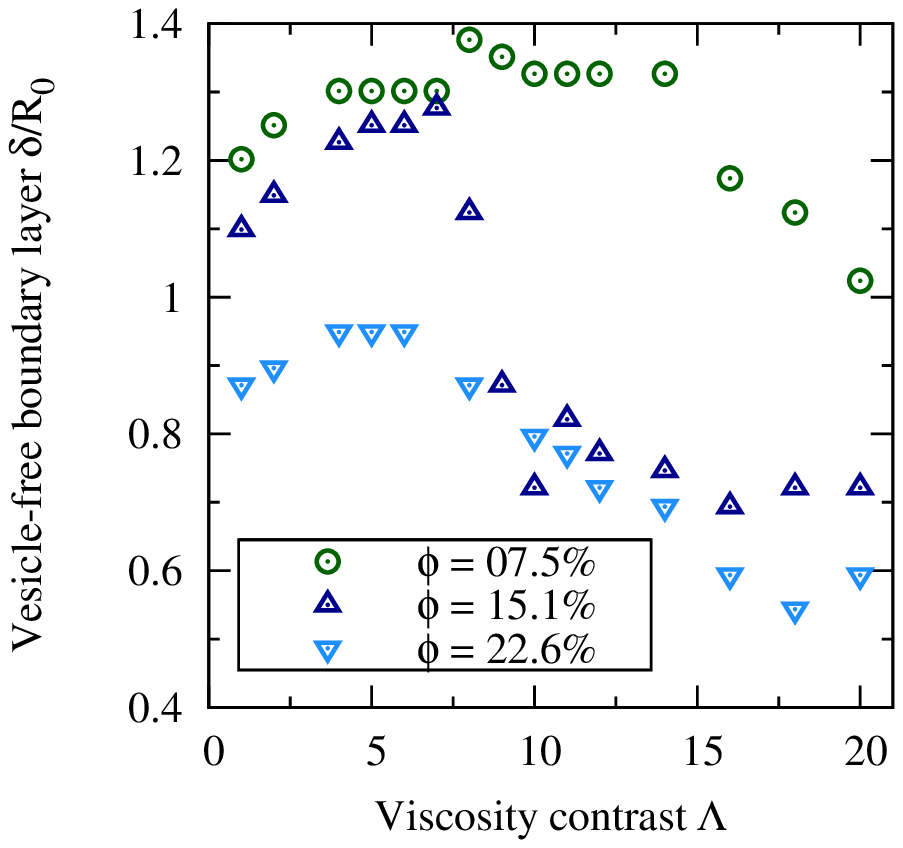}}
\caption{\label{fig:Figure09}
The rescaled vesicle-free boundary layer $\delta/R_0$ versus the
viscosity contrast $\Lambda$ of three suspensions with concentrations $\phi =
7.5\%$, $15.1\%$ and $22.6\%$. For all three concentrations, $\delta/R_0$
increases for low $\Lambda$ and then decreases for high $\Lambda$. Its maximum
coincides neither with the critical viscosity contrast of the
tank-treading-to-tumbling transition nor with the minimum of the intrinsic
viscosity. It does shift towards lower values of $\Lambda$ with increasing
$\phi$.  Other parameters: ${\rm Re}=0.5$, ${\rm Ca}=10$, $\Delta = 0.8$ and
$\chi = 0.2$.}
\end{figure}
In Fig.~\ref{fig:Figure06} we report the intrinsic viscosity $\eta$ versus the
viscosity contrast $\Lambda$ for three different concentrations: $\phi = 7.5\%,
15.1\%$ and $22.6\%$. Further, we provide the data corresponding to the case of a
single vesicle ($\phi=2.5\%$) as a reference. We observe that by increasing
the concentration, the intrinsic viscosity increases. The
curve of $\eta$ versus $\Lambda$ shifts towards higher values of $\eta$.
The more vesicles we have, the higher is the chance for vesicle-vesicle and
vesicle-wall collisions, which leads to an increase of the flow resistance. By
increasing $\phi$, $\Lambda$ at which $\eta$ reaches a minimum changes
the value without a clear trend. This is not consistent with the observation in
Ref.~\cite{Zhao2013} stating that the minimum of $\eta$ occurs at higher
$\Lambda$ with increasing $\phi$. However, Fig.~1 in Ref.~\cite{Vitkova2008}
clearly demonstrates that the minimum of the viscosity of red blood cell
suspensions occurs at lower $\Lambda$ with increasing $\phi$. The
non-increasing behavior of $\eta$ as a function of $\phi$ observed here and in
Ref.~\cite{Vitkova2008} maybe attributed to the low swelling degree of our
vesicles $\Delta=0.8$ and of the red blood cells $\Delta=0.65$, in contrast to
the vesicles used in Ref.~\cite{Zhao2013} which have $\Delta = 0.90$. 

The transition from decreasing to increasing behavior of $\eta$ as a function
of $\Lambda$ becomes less pronounced for larger $\phi$. If we extrapolate
the trend of $\eta$ versus $\Lambda$ to higher values of $\phi$ (the limit of
dense suspensions) we expect $\eta$ to be an almost constant function that does
not depend on $\Lambda$. At a higher concentration ($\phi = 22.6\%$), we observe
a continuous cross-over from vesicles with lower to the ones with higher
viscosity contrast. For denser suspensions, the tumbling is inhibited because
of the vesicle-vesicle and vesicle-wall hydrodynamic interactions (collisions)
that become more frequent. In this limit, the details of the dynamics of each
individual vesicle are irrelevant to the rheology of vesicle suspensions. 

For each concentration, we also measure the thickness of the vesicle-free boundary layer $\delta$ defined as the thickness of the gap that develops between the wall and the core vesicle-rich region of a suspension. The vesicle-free boundary layer is due to the wall-induced lift force that pushes vesicles away from the wall and results in their complete absence close to the wall. In Fig.~\ref{fig:Figure09} we report $\delta$ (averaged in time and scaled by the vesicle size) versus the viscosity contrast $\Lambda$ for three suspensions with concentrations $\phi = 7.5\%$, $15.1\%$
and $22.6\%$. For these three concentrations, $\delta/R_0$ is a non-monotonic function of $\Lambda$. It
increases for low $\Lambda$ and then decreases for high $\Lambda$. Its maximum
increases with increasing $\phi$ and it does shift towards lower values of
$\Lambda$. The maximum of $\delta$ coincides neither with the critical
viscosity contrast of the tank-treading-to-tumbling transition nor with the
minimum of the intrinsic viscosity. This means that the intrinsic viscosity and
the vesicle-free layer thickness are not correlated. Therefore, the
non-monotonic behavior of $\eta$ with $\Lambda$, which we capture in our study, is
tightly correlated with the vesicle dynamics and not with the vesicle-free
boundary layer thickness as is the case in Ref.~\cite{Lamura2013}.
\subsection{Effect of deformability ${\rm Ca}$}
\begin{figure}
\resizebox{\columnwidth}{!}{\includegraphics{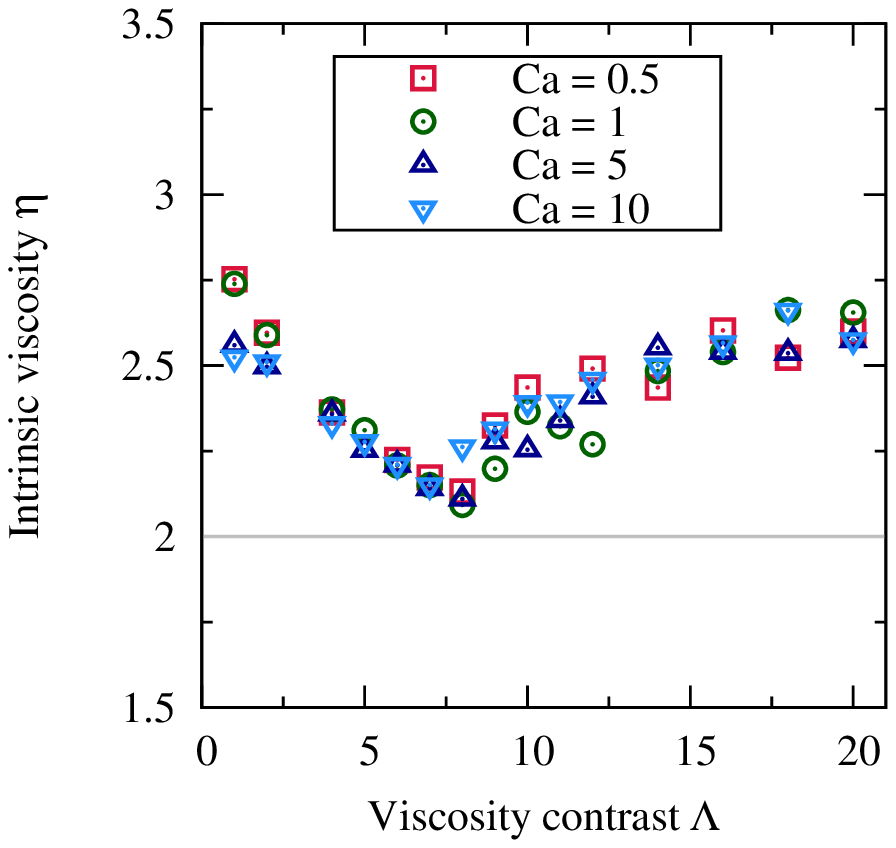}}
\caption{\label{fig:Figure07}
The intrinsic viscosity $\eta$ versus the viscosity contrast $\Lambda$ of four
suspensions with different capillary numbers ${\rm Ca}$: $0.05$, $1$, $5$ and
$10$. The deformability of the vesicles does not have any notable effect on the
macroscopic viscosity of the suspension. Other parameters: ${\rm Re}=0.5$, $\Delta = 0.8$, $\chi = 0.2$ and $\phi=15.1\%$.}
\end{figure}
The capillary number, ${\rm Ca}=\eta_0 \gamma R_0^3/\kappa$, controls
how the shape of a vesicle deforms in response to an applied external flow.
Vesicles deform less when ${\rm Ca} \ll 1$ (limit of stiffer vesicles) and
undergo larger deformation when ${\rm Ca} \gg 1$. We vary ${\rm Ca}$
by varying only the membrane rigidity $\kappa$. In this
way, all other parameters and in particular the shear rate $\gamma$ are hold
constant assuring also a constant Reynolds number (${\rm Re}=0.5$). 

In Fig.~\ref{fig:Figure07}, we report the intrinsic viscosity $\eta$ versus the
viscosity contrast $\Lambda$, for suspensions with concentration $\phi=15.1\%$ and with different capillary numbers: ${\rm Ca} = 0.5$, $1$, $5$ and
$10$. It appears that the vesicle deformability (${\rm Ca}$) does not have any
substantial effect on the viscosity of a suspension. $\eta$ still varies in a non-monotonic
way with $\Lambda$, but without any significant quantitative change when
varying ${\rm Ca}$. 
This is consistent with the results of
Refs.~\cite{Ghigliotti2010,Zhao2013}. Stiffer vesicles assume almost a similar
steady inclination angle as deformable vesicles when tank-treading. For
deformable vesicles, they tumble in an almost similar manner as stiffer
vesicles; their tumbling period is less affected by the shape deformability.
${\rm Ca}$ affects the shape deformation but not as much the dynamics. Moreover, in 2D
simulations, perfectly inextensible vesicles (the perimeter is kept
constant) do not exhibit vacillating-breathing motion (trembling) observed
theoretically and numerically for their 3D counterparts. The absence of this
dynamical state of motion in 2D (direct transition from TT to TB even at higher
${\rm Ca}$) is a further explanation of why $\eta$ is insensitive to variations in ${\rm Ca}$.
That $\eta$ does not depend on ${\rm Ca}$ means that the vesicle suspensions we study behave
like a Newtonian fluid. A ${\rm Ca}$-induced dynamical transition or ${\rm
Ca}$-induced variation of the free-vesicle boundary thickness are expected to
lead to non-Newtonian behavior. For example, the vesicle suspensions studied in
Ref.~\cite{Lamura2013} are non-Newtonian fluids exhibiting shear-thinning
behavior.
\subsection{Effect of the swelling degree $\Delta$}
\begin{figure}
\resizebox{\columnwidth}{!}{\includegraphics{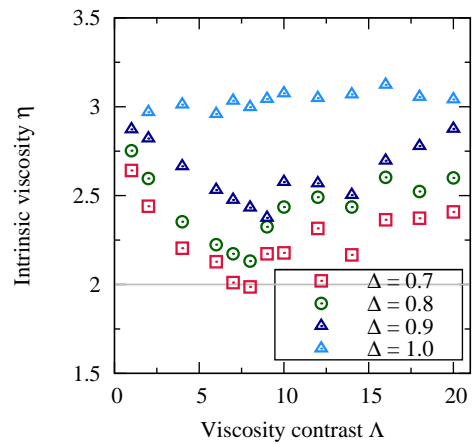}}
\caption{\label{fig:Figure08}
The intrinsic viscosity $\eta$ versus the viscosity contrast $\Lambda$ of four
suspensions with swelling degrees $\Delta = 0.7$, $0.8$, $0.9$ and
$1$. $\eta$ increases with $\Delta$ and its minimum shifts
towards higher values of $\Lambda$. For $\Delta = 1$, $\eta$ is a constant and
does not depend on $\Lambda$. Other parameters: ${\rm Re}=0.5$, ${\rm Ca =
0.5}$, and $\chi = 0.2$.}
\end{figure}
To investigate how the rheology of a suspension of vesicles is
affected by varying the swelling degree $\Delta$, we consider monodisperse
suspensions consisting of $6$ vesicles with size $R_0$. The swelling degree
$\Delta = 4\pi A/P^2$ is varied by swelling (deflating) vesicles, that is in
2D, by increasing (decreasing) $A$ while holding the perimeter $P$ of the
vesicles constant. This change in the enclosed area effectively leads to a
slight modification of the concentration: $\phi = 0.13 \%$ for $\Delta = 0.7$,
$\phi = 0.15 \%$ for $\Delta = 0.8$, $\phi = 0.17 \%$ for $\Delta = 0.9$ and
$\phi = 0.19 \%$ for $\Delta = 1$. All other parameters are similar to the
above sections, except for ${\rm Ca}$. Here, we opt for a smaller value (${\rm
Ca}=0.5$) in order to be able to capture the effect of vesicle shape (via
$\Delta$) while excluding the contribution of the shape deformation (induced by
larger ${\rm Ca}$).

In Fig.~\ref{fig:Figure08} the intrinsic viscosity $\eta$ versus the viscosity
contrast $\Lambda$ is reported  for various swelling degrees $\Delta$: $0.7$,
$0.8$, $0.9$ and $1$.  Again, we observe the same non-monotonic trend of $\eta$
as a function of $\Lambda$ with a minimum. Here, $\eta$ shifts upwards when
increasing $\Delta$. This is due to the increase of the flow resistance for
more swollen vesicles $\Delta \rightarrow 1$ (the limit of circular-shaped
vesicles in 2D). This observation agrees perfectly with the rheology of a single
vesicle without viscosity contrast ($\Lambda = 1$): $\eta$ is an increasing
function of $\Delta$~\cite{Ghigliotti2009}. This was explained by the fact that
the steady inclination of a tank-treading vesicle increases with $\Delta$.
However, for tumbling vesicles with viscosity contrast (see
Fig.~\ref{fig:Figure08}), the increase in $\eta$ for larger $\Delta$ maybe
attributed to the high tumbling frequency of less deflated vesicles, which
favors rotation of vesicles and thus decreases the flow resistance.
Moreover, we observe that the point of the minimum shifts to the right for
larger $\Delta$: deflated vesicles are more subject to tumbling motion than
swollen vesicles. This agrees with the fact that the critical viscosity for a
single vesicle increases with $\Delta$~\cite{Beaucourt2004}. For $\Delta = 1$
(circular vesicles in 2D), vesicles behave like rigid (circular) particles for
which TT or TB states are meaningless, therefore, $\eta$ does not exhibit the
non-monotonic behavior with $\Lambda$. Instead, it assumes a constant value
that is larger than the Einstein coefficient ($\eta = 2$), because of the
influence of confinement and interaction between particles.
\section{Discussions and conclusions}
In this article, we presented numerical simulations of the rheological behavior
of vesicle suspensions under shear flow as a function of the viscosity contrast
(the ratio between the viscosities of the encapsulated and the suspending
fluids). Our two-dimensional fluid-structure simulations are based on a
combination of the lattice-Boltzmann and the immersed boundary methods. The
method has been benchmarked against previous works performed for the case of a
single isolated vesicle suspended in unbounded creeping
flow~\cite{Danker2007,Ghigliotti2010}. As those authors, we recover the
non-monotonic behavior of the intrinsic viscosity versus the viscosity contrast
-- even in the presence of bounding walls and at non-zero Reynold number. In
contrast to a recent work by Lamura and Gompper~\cite{Lamura2013} we found that
the effect proposed by Danker and Misbah~\cite{Danker2007} persists even for
non-dilute suspensions of vesicles and when we vary the deformability and the
swelling degree of the vesicles. The effect becomes less pronounced at higher
swelling degrees and at higher concentrations (limit of dense suspensions) where
the tumbling motion is inhibited. This can be understood by means of the
vesicle-vesicle and vesicle-wall hydrodynamic collisions that then become more
important.

Let us now close the question about the origin of the apparent contradicting
behaviors of the intrinsic viscosity versus the viscosity contrast ($1 \leq \Lambda \leq 10$)
reported in different studies~\cite{Danker2007,Vitkova2008,Ghigliotti2010,Zhao2011,Zhao2013,Thiebaud2013,Lamura2013,Kantsler2008}:
If we disregard errors in the measurements, numerical artifacts or the
contribution of thermal fluctuations, the
influence of the wall confinement remains the main possible origin. 
Weak confinements, for example, $\chi = 0$ in Refs.~\cite{Ghigliotti2010,Zhao2013} or $\chi = 0.2$ in the present work, allow for the tank-treading-to-tumbling transition to take place. 
This dynamical transition, triggered solely by increasing the viscosity contrast, is the main responsible mechanism for the non-monotonic behavior of the intrinsic viscosity we observe.
Increasing confinement delays the transition to the tumbling
motion~\cite{Kaoui2012}, as for example, in cone-plate
rheometers~\cite{Fischer2013}. This explains why in Ref.~\cite{Kantsler2008}
the intrinsic viscosity does not show an increasing behavior. The minimum (the
transition point) is in fact expected to occur at higher values of the
viscosity contrast. The authors even mentioned that their last data point,
taken at $\Lambda=10$, deviates from the decreasing monotonic behavior of the
viscosity. For these experiments, one would capture the minimum and the
increasing behavior of the intrinsic viscosity just by further increasing the
viscosity contrast beyond $\Lambda=10$.
However, the data in Ref.~\cite{Lamura2013} are obtained at higher
degrees of confinement ($\chi = 0.30$ and $0.35$). The walls are so close that
they strongly influence the dynamics and the microstructures formed by the vesicles. In
our case, such higher confinements do not even allow for the
tank-treading-to-tumbling transition to take place at $\Lambda <
7.8$~\cite{Kaoui2012}. In Ref.~\cite{Lamura2013} the interplay between the wall-induced lift force
and increasing the viscosity contrast causes the vesicle-free boundary layer to become narrower, and results in higher intrinsic viscosity. 
This is similar
to the F\aa hr\ae us-Lindqvist effect~\cite{Fahraeus1931}. Therefore, the
observed monotonic increasing behavior of the intrinsic visocity with the
viscosity contrast in Ref.~\cite{Lamura2013} is mainly due to the variation of
the vesicle-free boundary layer thickness, and not due to the vesicle dynamical
transition as is the case for the Danker-Misbah effect. For these simulations,
the wall effects are dominant and hide the contribution of the vesicle
dynamical transition. Thus, in order to capture the Danker-Misbah effect one
should decrease confinement to lower values, as is done in the present work. 
\section*{Acknowledgments}
We thank Chaouqi Misbah and the anonymous referees for valuable comments, 
and NWO/STW for financial support (VIDI grant 10787 of J. Harting).
\end{document}